\documentclass[%
  reprint,
groupedaddress,
 amsmath,amssymb,
 aps,
]{revtex4-1}

\usepackage{graphicx}

\usepackage{amsmath}
\usepackage{dcolumn}
\usepackage{bm}
\usepackage{comment}
\usepackage{braket}
\usepackage[mathlines]{lineno}

\newcommand{\tTr}{\mathrm{tTr}}

\begin{document}


\title{Boundary Tensor Renormalization Group}


\author{Shumpei Iino}
\email[E-mail address: ]{iino@issp.u-tokyo.ac.jp}
\affiliation{Institute for Solid State Physics, The University of Tokyo, Kashiwa, Chiba, Japan}
\author{Satoshi Morita}
\affiliation{Institute for Solid State Physics, The University of Tokyo, Kashiwa, Chiba, Japan}
\author{Naoki Kawashima}
\affiliation{Institute for Solid State Physics, The University of Tokyo, Kashiwa, Chiba, Japan}


\date{\today}

\begin{abstract}
  We develop the tensor renormalization group (TRG) algorithm for statistical systems with open boundaries, which allows us to investigate not only the bulk but also the boundary property, such as the surface magnetization. We demonstrate that the tensors representing the boundary in our algorithm exhibit the fixed point structures just as bulk tensors in previous TRG algorithms. At criticality, the scale-invariant boundary fixed point tensors have the information of the conformal tower, which is described by the underlying boundary conformal field theory.
\end{abstract}


\maketitle



\section{Introduction\label{sec:intro}}

Renormalization group (RG) is one of the most significant concepts in modern physics~\cite{Wilson1971-1,Wilson1971-2}. Apart from its original motivation, the prescription for the divergent physical quantity in the quantum field theory, the RG method has been useful to classify the phases, investigate the critical phenomena, and so on~\cite{Cardy_statmech}. The philosophy of RG has also been adopted to invent efficient numerical methods, such as density matrix renormalization group (DMRG)~\cite{White1992,White1993}, corner transfer matrix renormalization group (CTMRG)~\cite{Nishino1996,Nishino1997}, and entanglement renormalization~\cite{Vidal2007}. Recent interpretation of RG, `efficient compression of information' draws attention in the field of information science~\cite{PRL.116.140403}, especially machine learning~\cite{Mehta2014}.

Combining the real-space RG concept with tensor network representation of the partition function, Levin and Nave proposed tensor renormalization group (TRG) algorithm to contract the tensor network of the Boltzmann weight for statistical systems~\cite{Levin2007}. In addition to the capability to compute free energy with high accuracy, as they pointed out, the renormalized tensors show fixed point structures characterizing the corresponding phases. Gu and Wen investigated the precise meaning of this statement, and clarified that it consists of trivial tensors~\cite{Gu2009}. Further interestingly, the fixed point tensor at a critical point becomes scale invariant, from which one can extract the information of the underlying conformal field theory (CFT).

`Boundary' is another significant keyword in modern physics. The remarkable feature of topological insulators is the metallic surface state even though the bulk is insulator, and Majorana fermions emerge at the edge of topological superconductors~\cite{Hasan2010,Qi2011}. The symmetry protected topological (SPT) phases have the gapless or degenerated nontrivial surface states, which cannot be broken down by perturbation conserving the corresponding symmetries~\cite{Gu2009,Chen2012}. Even before the emergence of these recent `topological' topics, boundary physics and surface critical phenomena have been traditionally important subject of study~\cite{Binder1983_DG}. At the bulk critical point, the diverging correlation length of the bulk induces the singularity at the surface, which results in the different critical exponents of the surface physical quantities from the bulk ones (which is called extraordinary or ordinary transition depending on whether the surface was already ordered or not before the bulk transition). When the surface itself is also critical, another surface universality can emerge (called special transition). Especially for two-dimensional systems with one-dimensional edges, some of those exponents can be exactly calculated using the boundary conformal field theory (BCFT)~\cite{Cardy1987_DG}. Combining the above topics, there are recent attempts to study novel surface criticality in the quantum phase transition of the SPT phases~\cite{Zhang2017}.

Though there have been proposed many improved TRG algorithms after the invention of it~\cite{Gu2009,Xie2009,Zhao2010,Xie2012,Evenbly2015,Yang2017,Bal2017,Hauru2018,PRE.97.033310,PRB.99.155101}, very few studies by TRG-type tensor network methods have focused on the effect of boundaries or the physics arising in boundaries. This might be because generally in tensor network computation one can easily achieve a huge system size or deal with an infinite system by imposing the translational invariance on the tensors.

In contrast to most of previous TRG-type calculations that assume implicitly periodic boundary condition, in this paper, we use open boundaries and investigate a natural generalization of the higher order TRG (HOTRG)~\cite{Xie2012} algorithm so as to simulate the boundary effects. We call this algorithm \textit{boundary tensor renormalization group} (BTRG) below. As we shall demonstrate using the two-dimensional Ising model, BTRG allows us to compute the surface property such as the surface magnetization with the same computational complexity as HOTRG. In addition, the boundary tensors in our algorithm also converge to the fixed point tensors as the conventional TRG algorithms, which are the trivial tensors for disordered phase and the direct sum of them for the ordered phase. The fixed point tensor at the critical point has the conformal data reflecting the operator content of the corresponding BCFT.

In the next section, we construct the algorithm of BTRG, and describe how to analyze the fixed point boundary tensors. In Sec.~\ref{sec:result}, the benchmark computation of the two-dimensional Ising model is shown, and the conclusion is in Sec.~\ref{sec:conclusion}. In the appendix, we explain how to compute a proper projector for the renormalization and how to obtain the scale invariant fixed point tensor at criticality.
%

\section{Boundary Tensor Renormalization Group\label{sec:algorithm}}

In this paper, we consider the two-dimensional square lattice on an open cylinder, where we adopt periodic boundary condition for one direction of the two while the open boundary condition for the other, as shown in Fig.~\ref{fig:RGstep} (a). Just as HOTRG, our algorithm can be easily generalized for the $d$-dimensional hyper-cubic lattice. For further simplicity, we assume the tensor network is translationally invariant along the periodic direction.
%

\subsection{Algorithm}
\begin{figure}
\includegraphics[width=6cm]{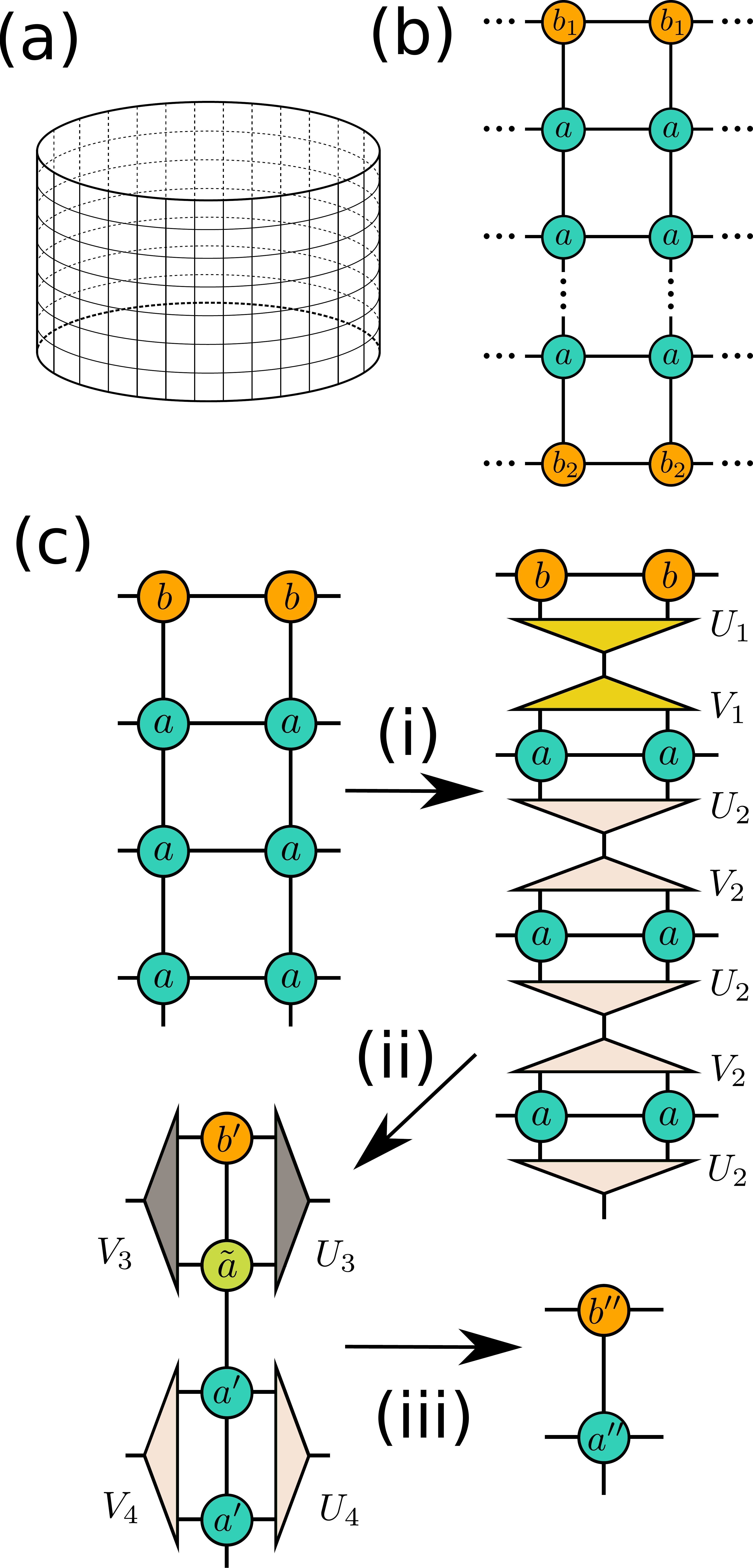}%
\caption{\label{fig:RGstep}
(Color online) (a) The square lattice on an open cylinder. (b) The tensor network of the annulus geometry, with three type tensors, the bulk tensor $a$, and the boundary tensors $b_1$ and $b_2$. (c) One renormalization step of the BTRG algorithm. (i) The projectors are inserted into every two vertical bonds to contract two neighbouring tensors. (ii) After the horizontal contraction, the projectors are inserted into every two horizontal bonds. (iii) Updating the horizontal bond, we come back to the initial network with a quarter of the previous system size.}
\end{figure}

In BTRG algorithm, we hold three tensors: a rank-four bulk tensor $a$ and two rank-three boundary tensors $b_1$ and $b_2$ which respectively represent the two open edges, see Fig.~\ref{fig:RGstep} (b). In Fig.~\ref{fig:RGstep} (c), the procedure of one RG step is graphically shown for the upper edge of the cylinder. First, every two tensors horizontally neighbouring is renormalized into one tensor. In the step (i) in Fig.~\ref{fig:RGstep} (c), the projector $U$ and $V$ are inserted into every two vertical bonds, which can be created using the four tensors connected by the two bonds. For instance, the projectors $U_1$ and $V_1$ inserted between the boundary tensors $b$ and the nearest bulk tensors $a$ are determined so as to minimize the following cost function $\mathcal{C}$ keeping the maximal bond dimension for truncation lower than some threshold $\chi$:
\begin{equation}
  \mathcal{C}=\left|
  \begin{minipage}{1.5truecm}
      \centering
      \includegraphics[width=1.5truecm,clip]{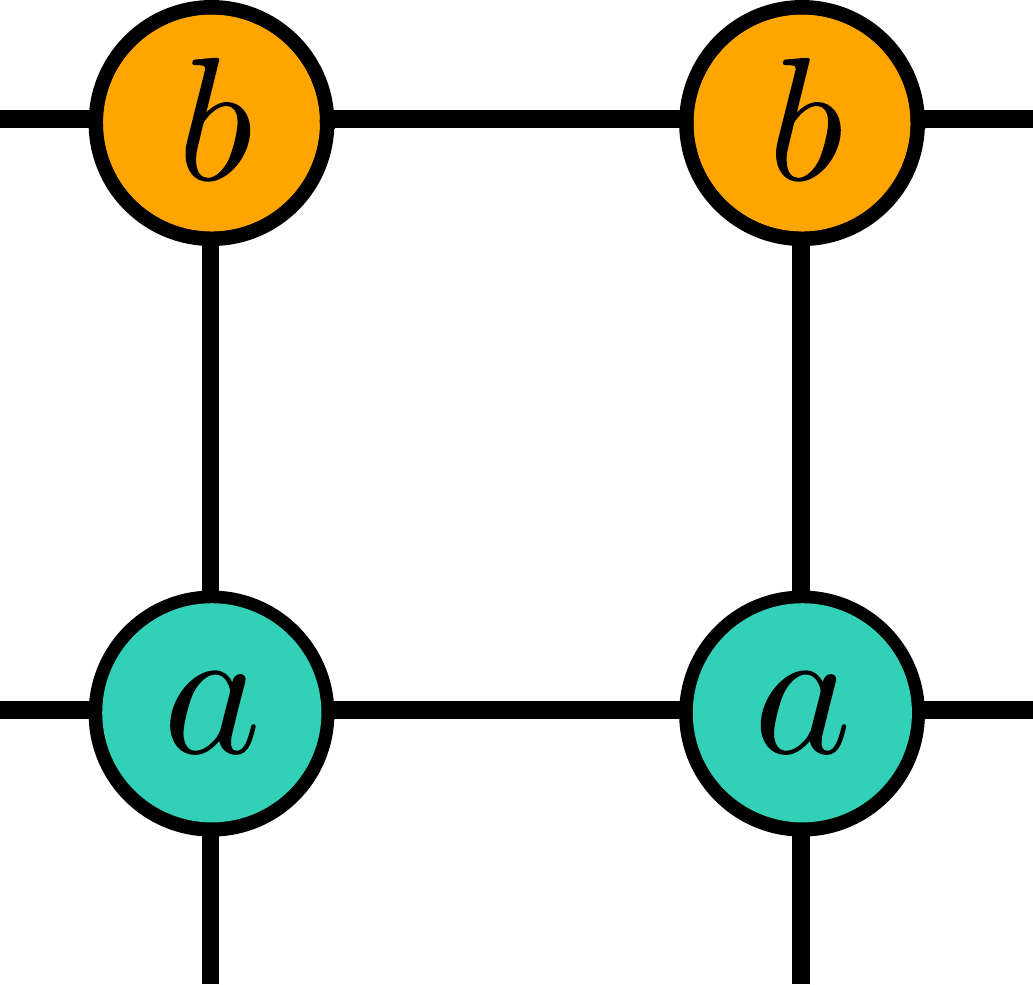}
\end{minipage}
-
\begin{minipage}{1.65truecm}
      \centering
      \includegraphics[width=1.65truecm,clip]{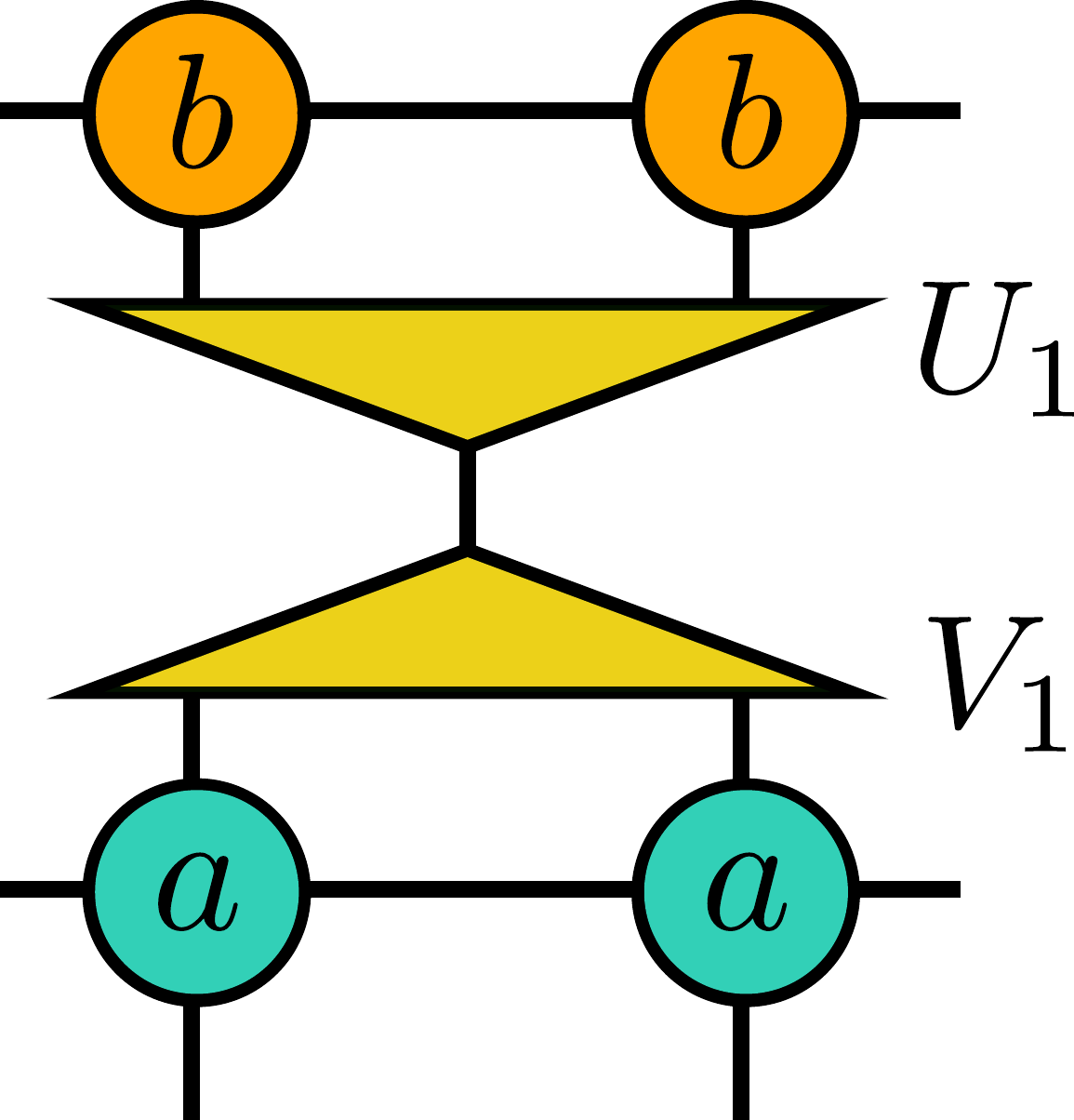}
\end{minipage}
\right|^2.
\label{eq:costfunc_ex}
\end{equation}
We can obtain the projectors without calculating the norm directly, as explained in the appendix~\ref{sec:projector}. The projectors $U_2$ and $V_2$ are generated only from the bulk tensors in the same way and inserted into them. In the next step (ii), after we contract the projectors and horizontal pair of tensors, the projectors for the vertical contraction are created. Notice that, we have an intermediate tensor $\tilde{a}:=V_1(aa)U_2$ vertically next to the boundary tensor, which is different from the $a':=V_2(aa)U_2$. Just as Eq.~(\ref{eq:costfunc_ex}), we can generate two pair of projectors for renormalizing $b'$ and $\tilde{a}$ and two bulk tensors. After the contraction (iii), one step of the real-space RG with the scale factor 2 is completed. The RG for the other side of boundary can be performed in the same way. If the system is finite, after repeating this step a number of times, the contraction of all the network can be computed as the trace of two boundary tensors: e.g., the partition function for a $2^t\times 2^{t+1}$ system is calculated as
\begin{equation}
  Z=
\begin{minipage}{0.75truecm}
      \centering
      \includegraphics[width=0.75truecm,clip]{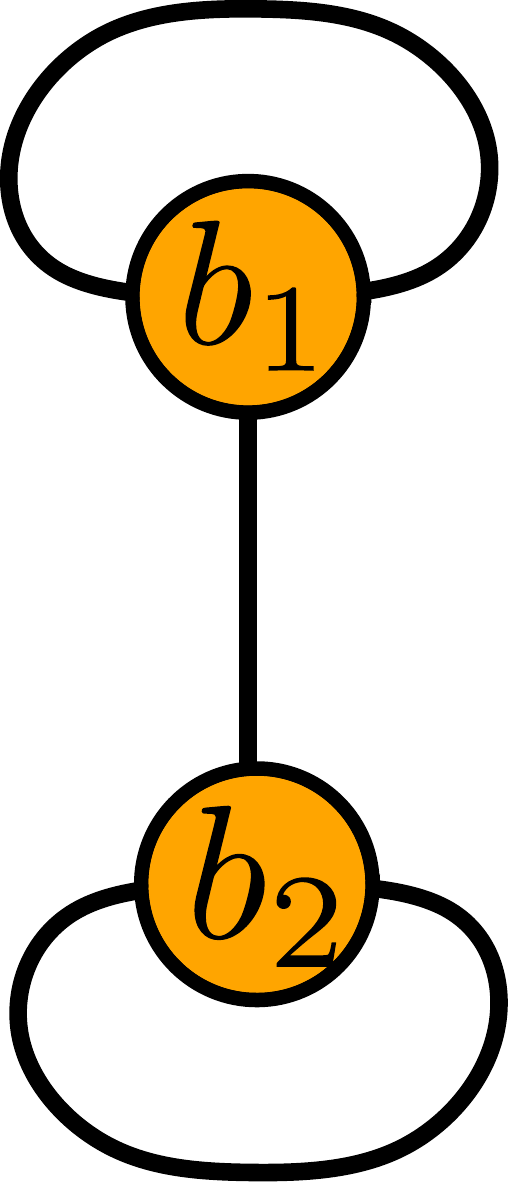}
\end{minipage},
\end{equation}
after $t$-th RG steps. The computational complexity is the same as HOTRG algorithm, $O(\chi^7)$ for two-dimensional case.
%

\subsection{Fixed point tensor analysis\label{subsec:fptensor}}

As Levin and Nave pointed out~\cite{Levin2007}, after enough RG steps all the tensors converge to fixed point tensors which characterize what phases the system is in. Gu and Wen developed the theory of fixed point tensors in Ref.~\onlinecite{Gu2009}: they clarified that ideally the fixed point tensor for a trivial phase without symmetry breaking or long-range entanglement is a trivial tensor $T^{\mathrm{TRI}}$, all of whose bond dimensions are one, and for symmetry broken phases it is the direct sum of the same number of $T^{\mathrm{TRI}}$ as the degeneracy of the phase. For instance, the fixed point tensor for the ordered phase of the Ising model is
\begin{equation}
  T = T^{\mathrm{TRI}}\oplus T^{\mathrm{TRI}}.
  \label{eq:fptensorZ2}
\end{equation}
Also, they proposed a method of detecting symmetry breaking phase transitions utilizing this property. The following quantity $X$ for the bulk tensor $a$ is a step function, whose value of symmetry broken phases is equal to the degeneracy while one for the trivial phase:
\begin{equation}
  X = \frac{\left[\tTr\left(
    \begin{minipage}{1.0truecm}
      \centering
      \includegraphics[width=1.0truecm,clip]{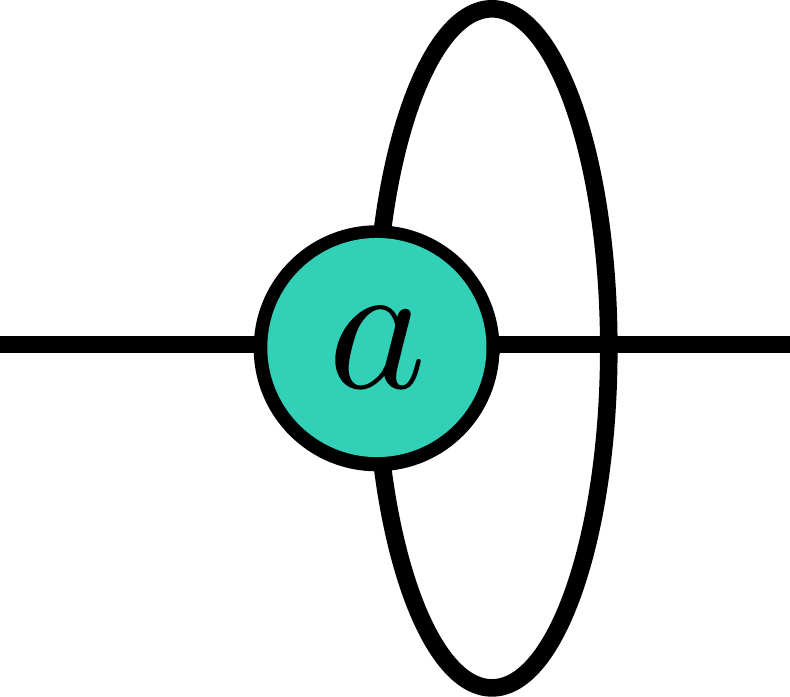}
    \end{minipage}
  \right)\right]^2}
  {
  \tTr\left[\left(
    \begin{minipage}{1.0truecm}
      \centering
      \includegraphics[width=1.0truecm,clip]{trfmat-a.pdf}
    \end{minipage}
  \right)^2\right]}.
  \label{eq:Xbulk}
\end{equation}
It can be confirmed in actual simulation of the Ising model that the eigenvalue spectrum of the transfer matrix in Eq.~(\ref{eq:Xbulk}) is almost zero except the largest one for the trivial phase, whereas for the ordered phases also zero except the largest two eigenvalues, which results in the step-function feature of $X$.

Similarly, the boundary tensors for BTRG algorithm are also renormalized into fixed point tensors, as demonstrated in the next section. The fixed point tensor for the trivial phases is a rank-three tensor whose bonds are all one-dimensional, and also the direct sum of it for symmetry broken phases. The phase-transition detector for boundary tensors can be defined as
\begin{equation}
  X_{\textrm{s}} = \frac{\left[\tTr\left(
    \begin{minipage}{0.6truecm}
      \centering
      \includegraphics[width=0.6truecm,clip]{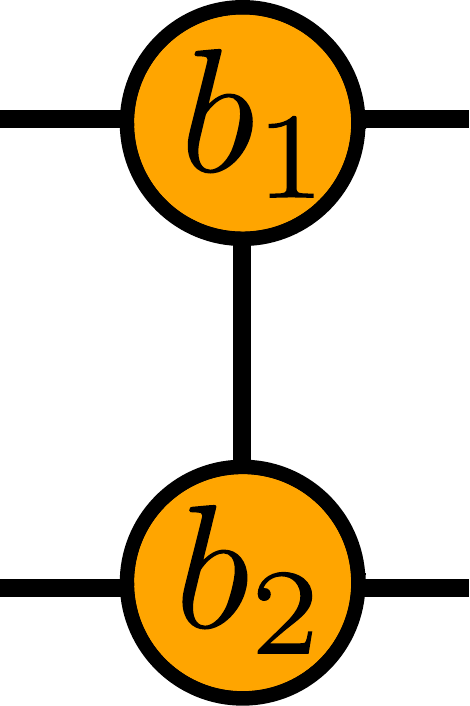}
    \end{minipage}
  \right)\right]^2}
  {
  \tTr\left[\left(
    \begin{minipage}{0.6truecm}
      \centering
      \includegraphics[width=0.6truecm,clip]{trfmat-b.pdf}
    \end{minipage}
  \right)^2\right]}.
  \label{eq:Xsurf}
\end{equation}
This quantity shows the same behavior as Eq.~(\ref{eq:Xbulk}), as confirmed in the next section.

If the system is at criticality, tensors are renormalized into some infinite dimensional ones different from $T^{\mathrm{TRI}}$. In this case, analyzing conformal invariance of the fixed point tensors makes it possible to obtain conformal data described by the corresponding CFT. The analysis of the bulk tensors is in detail described in the appendix of Ref.~\onlinecite{Gu2009}. We can derive similarly how to extract the conformal data from the boundary tensors: BCFT yields the partition function in the annulus geometry with $M$ height and $L$ circumference (see Fig.~\ref{fig:RGstep} (a)) is~\cite{Francesco_CFT}
\begin{equation}
  Z = \mathrm{Tr} \exp\left[-\frac{M}{L}\pi\left(\hat{L}_0-\frac{c}{24}\right) \right],
  \label{eq:Z_BCFT}
\end{equation}
where $c$ is the central charge and $\hat{L}_0$ is the Virasoro operator whose eigenstates are the primary fields and their descendants. In tensor network representation, since the scale invariant term in partition function is described by the trace of the scale invariant tensors (see Ref.~\onlinecite{Gu2009}), the formula Eq.~(\ref{eq:Z_BCFT}) can be applied for the network constructed by the scale invariant tensors. As for the way to obtain scale invariant tensors, see the appendix~\ref{sec:invtensor}. For example, if we choose $L=2$ and construct a transfer matrix $B$ from two scale-invariant boundary tensors $b_{1\mathrm{inv}}$ and $b_{2\mathrm{inv}}$,
\begin{eqnarray}
  Z&=&\tTr\left[
\begin{minipage}{1.2truecm}
      \centering
      \includegraphics[width=1.2truecm,clip]{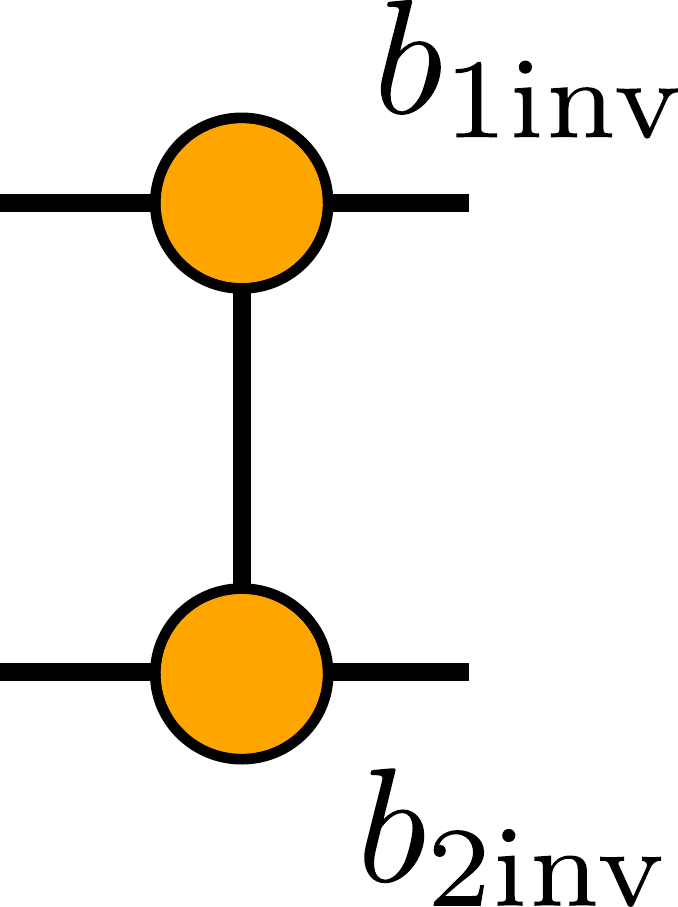}
\end{minipage}
\right]^M \equiv \tTr\left[
\begin{minipage}{1.0truecm}
      \centering
      \includegraphics[width=1.0truecm,clip]{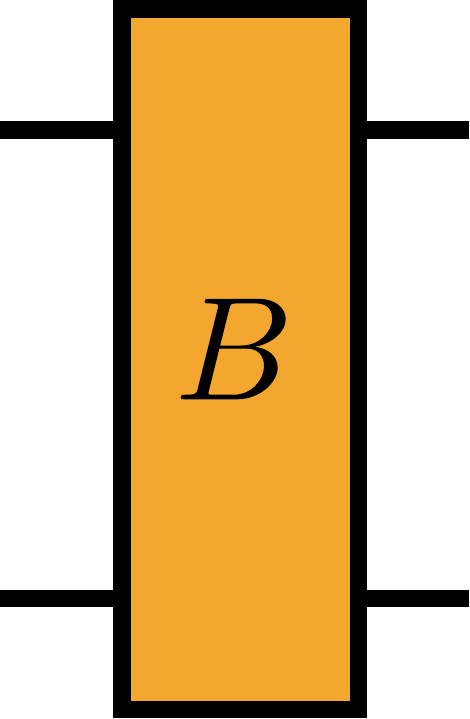}
\end{minipage}
\right]^M
\label{eq:trfmat_bb}\\
  &=& \mathrm{Tr} \exp\left[-\frac{M}{2}\pi\left(\hat{L}_0-\frac{c}{24}\right) \right].
\label{eq:trfmat_B}
\end{eqnarray}
If we describe the eigenvalue spectrum of $B$ as $\{\lambda_n\}$, the conformal dimensions $h_n$ can be computed as
\begin{equation}
  \ln\lambda_n = -\frac{\pi}{2}\left(h_n-\frac{c}{24}\right).
  \label{eq:BCFT_tower}
\end{equation}
If we use known values of the lowest conformal dimension $h_0$ or the central charge $c$, we can determine the whole conformal spectrum.

In the end of this section, we would like to notify that in order to obtain the correct fixed point tensor we have to eliminate the short entanglement loops, which remain after converging to fixed point and waste the capacity of the tensors~\cite{Gu2009}. The above explained BTRG algorithm as it is cannot remove the short correlation. Therefore, to achieve the fixed point tensor with high accuracy, it is necessary to combine with such an algorithm as entanglement filtering in loop-TNR~\cite{Yang2017}, GILT algorithm~\cite{Hauru2018}, entanglement branching~\cite{Harada2018}, or full environment truncation~\cite{PRB.98.085155}.
%

\section{Numerical Results\label{sec:result}}

To evaluate performance of our algorithm, we simulate the two-dimensional ferromagnetic Ising model~\cite{Potts1952}, whose Hamiltonian is
\begin{equation}
  \beta\mathcal{H}=-K\sum_{\langle ij\rangle\in\mathrm{bulk}}\delta_{\sigma_i\sigma_j}
  -K_s\sum_{\langle ij\rangle\in\mathrm{surface}}\delta_{\sigma_i\sigma_j},
  \label{eq:potts_hamiltonian}
\end{equation}
where $\delta$ is the Kronecker delta and $\sigma=-1$ or $+1$. If the both spins of $\sigma_i$ and $\sigma_j$ are on the edges, the nearest-neighbour coupling constant is $K_s$ otherwise $K$. The lattice geometry is the annulus geometry as depicted in Fig.~\ref{fig:RGstep} (a).

\subsection{Magnetization}

\begin{figure}
\includegraphics[width=7cm]{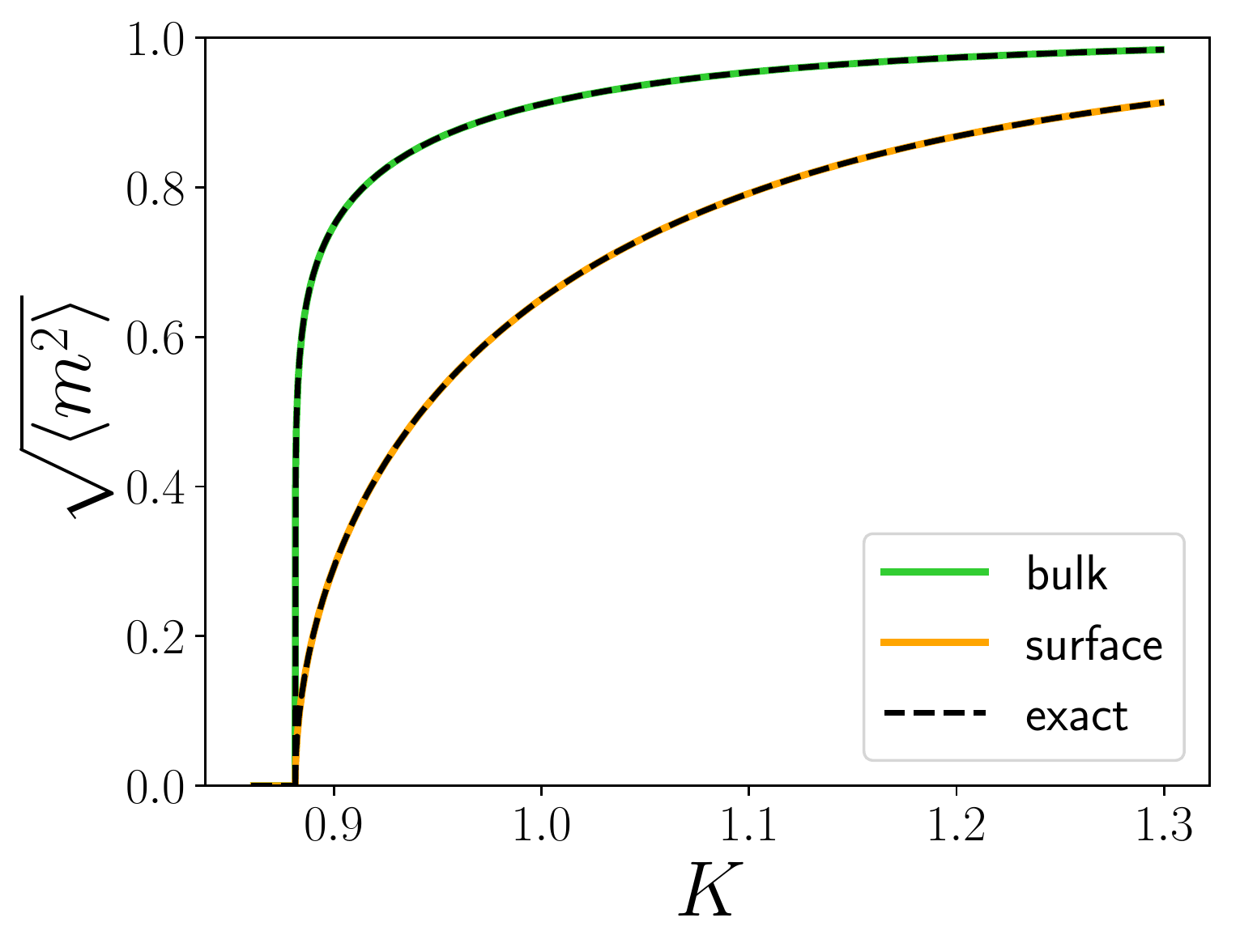}%
\caption{\label{fig:q2mag}
(Color online) The spontaneous magnetizations in the bulk and the surface computed by the BTRG algorithm for $\chi=16$. The dashed line shows the exact results~\cite{Yang1952,McCoy1967}.}
\end{figure}

Using the impurity tensor method (see, e.g., Ref.~\onlinecite{Morita2018}), we compute the spontaneous magnetization $m$ in the bulk and the surface, which are defined as
\begin{eqnarray}
  \mathrm{bulk:}\ \ &m_{\textrm{b}}& = \frac{1}{L^2}\sum_{i}\sigma_i\\
  \mathrm{surface:}\ \ &m_1& = \frac{1}{2L}\sum_{i\in\mathrm{surface}}\sigma_i,
  \label{eq:mdefinition}
\end{eqnarray}
where $L$ is the system size. As shown in Fig.~\ref{fig:q2mag}, compared with the exact results~\cite{Yang1952,McCoy1967}, the computed magnetizations for $K_s=K$ are quantitatively good even for such a small bond dimension as $\chi=16$. Notice that we employ $\sqrt{\langle m^2\rangle}$ as an order parameter since $\langle m\rangle$ is always zero for the $Z_2$ symmetric tensor~\cite{Singh2011}.

\subsection{Fixed point tensor in the non-critical phases}

\begin{figure}
\includegraphics[width=7cm]{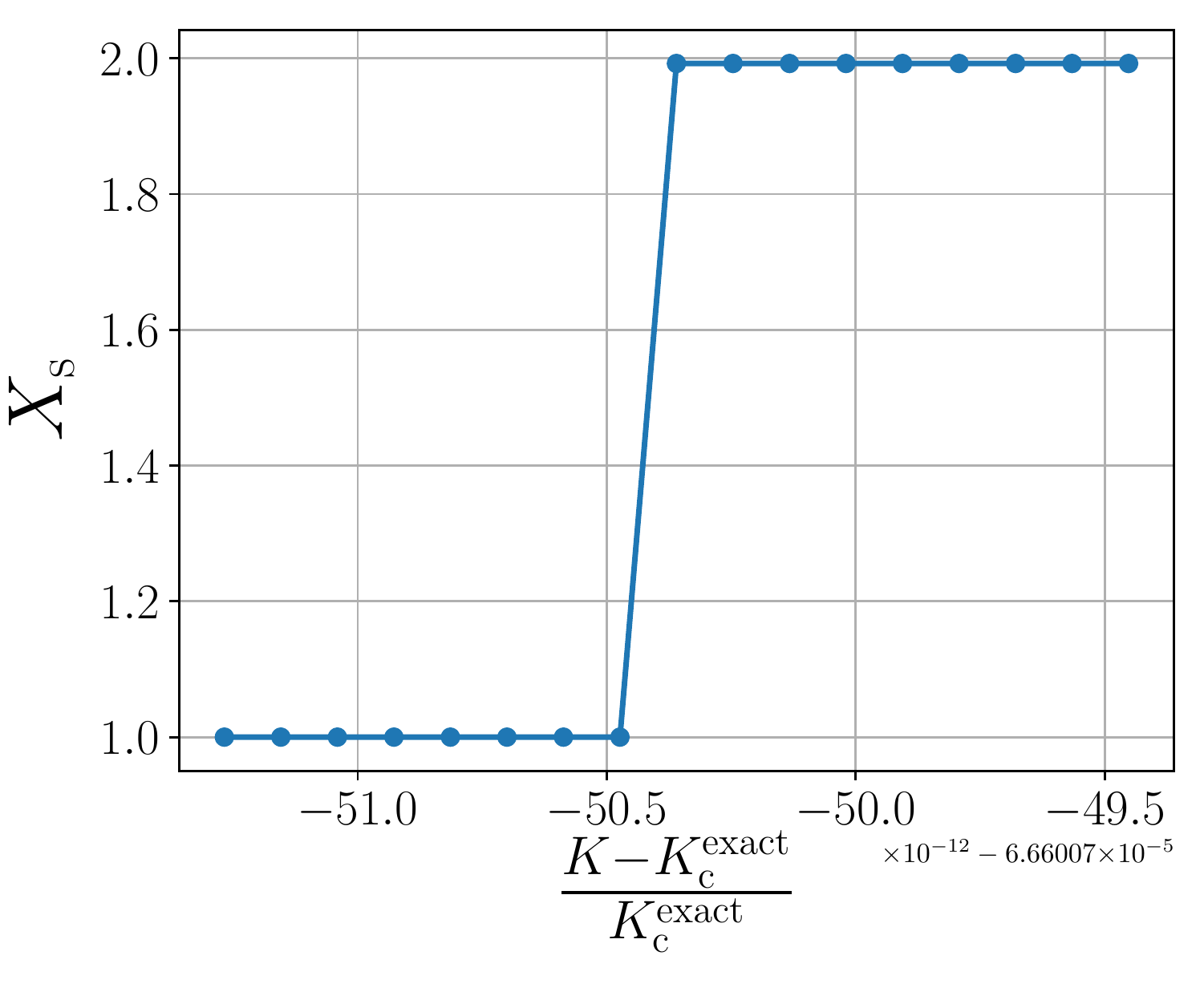}%
\caption{\label{fig:q2X}
(Color online) The function $X$ defined in Eq.~(\ref{eq:Xsurf}) computed for $\chi=16$ around the transition point. The exact critical point is $K_{\textrm{c}}^{\textrm{exact}}=\ln(1+\sqrt{2})$~\cite{Wu1982}.}
\end{figure}

We analyze the fixed point structure of the boundary tensors for non-critical phases. In Fig.~\ref{fig:q2X}, Eq.~(\ref{eq:Xsurf}) is computed until the convergence in the very narrow temperature region for $\chi=16$ and $K_s=K$. The values in the disordered phase and ordered phase are respectively one and two as expected, and we can estimate the transition point $K_{\textrm{c}}$ for $\chi=16$ as $K_{\textrm{c}}= 0.881314886877(2)$. The relative error from the exact critical temperature~\cite{Wu1982} $K_{\textrm{c}}^{\textrm{exact}}=\ln(1+\sqrt{2})$ is about $6.7\times 10^{-5}$, which is consistent with the transition point obtained from the crossing point of the Binder ratio for the same bond dimension in Ref.~\onlinecite{Morita2018}. Because the quantity defined Eq.~(\ref{eq:Xsurf}) reacts sharply for such a subtle change of temperature, the transition temperature for a given bond dimension can be determined with very high precision.

\subsection{Fixed point tensor at criticality}

We compute the central charge and the conformal towers from the boundary tensors, which corresponds to the minimal CFT $\mathcal{M}_{4,3}$ in the annulus geometry~\cite{Francesco_CFT}. From the bulk fixed point tensor of the Ising model, as already confirmed in many preceding works with periodic boundary condition~\cite{Gu2009,Evenbly2015,Yang2017,Bal2017,Hauru2018,Harada2018}, one can extract the conformal tower generated from three primary operators: $\phi_{0,0}$, $\phi_{1/16,1/16}$, $\phi_{1/2,1/2}$, where the subscripts represent the conformal dimensions for holomorphic and antiholomorphic part of the Virasoro algebra respectively. On the other hand, the existence of the boundary puts a constraint on the Virasoro algebra, and the operator content of the CFT changes according to what boundary conditions are imposed~\cite{CARDY1986200}.

Given the boundary conditions of both sides in the annulus, the operator content can be calculated easily by fusion rules~\cite{CARDY1989581}. Corresponding to the primary fields, there are three Cardy states (i.e., conformally invariant boundary conditions) in the Ising CFT:
\begin{eqnarray}
  \Ket{+} &=& \Ket{\tilde{0}}\label{eq:Cardy+}\\
  \Ket{-} &=& \Ket{\tilde{\frac{1}{2}}}\label{eq:Cardy-}\\
  \Ket{\mathrm{free}} &=& \Ket{\tilde{\frac{1}{16}}}.
\end{eqnarray}
$\Ket{+}$ and $\Ket{-}$ represent the fixed boundary states, where the spins in the edge are all $\sigma=+1$ and all $\sigma=-1$ respectively. In the Hamiltonian Eq.~(\ref{eq:potts_hamiltonian}), when $K_s=K$ the surface is disordered and we associate this with free boundary condition $\Ket{\mathrm{free}}$ below, while $K_s\rightarrow\infty$ leads to the spontaneously ordered surface state $\Ket{\mathrm{fixed}}=\left(\Ket{+}+\Ket{-}\right)$, which is called fixed boundary condition below. The operator content for the system sandwitched by two boundary states $\Bra{\tilde{a}}$ and $\Ket{\tilde{b}}$ is determined as the result of the fusion rule $\phi_{a}\times\phi_{b}$. Therefore the operator contents for those boundary states are, from the fusion rules of the Ising CFT,
\begin{eqnarray}
  Z_{\mathrm{free,free}} &=& \chi_{0}+\chi_{\frac{1}{2}}\label{eq:free-free}\\
  Z_{\mathrm{fixed,fixed}} &=& 2\left[\chi_{0}+\chi_{\frac{1}{2}}\right]\label{eq:fix-fix}\\
  Z_{\mathrm{free,fixed}} &=& 2\chi_{\frac{1}{16}}\label{eq:free-fix},
\end{eqnarray}
where $\chi_{h}$ is the Virasoro character of the Verma module with conformal dimension $h$. For example, the partition function Eq.~(\ref{eq:free-free}) results from the fusion rule $\phi_{1/16}\times\phi_{1/16}=\phi_0+\phi_{1/2}$, and Eq.~(\ref{eq:fix-fix}) can be obtained as
\begin{equation}
  \left[\phi_{0}+\phi_{\frac{1}{2}}\right]\times\left[\phi_{0}+\phi_{\frac{1}{2}}\right]=2\left[\phi_{0}+\phi_{\frac{1}{2}}\right].
\end{equation}
Similar formulae can be found in Ref.~\onlinecite{Affleck_2000}.

\begin{figure}
\includegraphics[width=7cm]{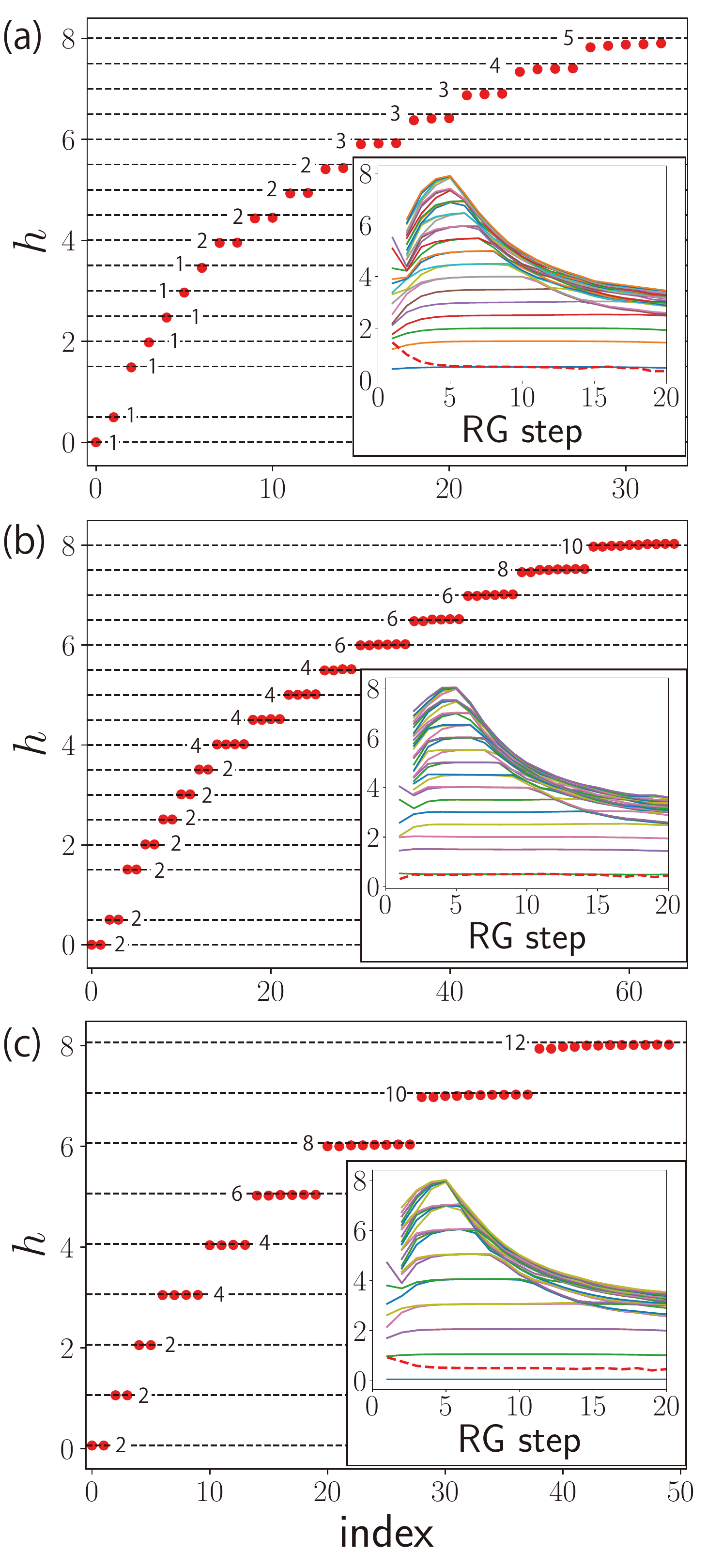}%
\caption{\label{fig:q2cfts}
(Color online) The conformal dimensions $h_n$ obtained at the fifth RG step of BTRG with $\chi=72$ for various boundary conditions. Note that the lowest one $h_0$ is given by hand. The exact values are denoted as the dashed line and the exact degeneracy is near the data plots. Each figure represents the case where (a) both edges are free boundary conditions, (b) both edges are fixed boundary conditions, and (c) the one edge is fixed while the other is free boundary condition. The operator contents for them correspond to Eq.~(\ref{eq:free-free}), Eq.~(\ref{eq:fix-fix}), and Eq.~(\ref{eq:free-fix}) respectively.  Inset: The flow of the central charge and the conformal dimensions for the RG step. The red dashed line represents that of the central charge.}
\end{figure}

To investigate the conformal fixed point tensor, we perform BTRG computation with $\chi=72$ at the exactly known critical point $K=K_{\textrm{c}}^{\textrm{exact}}$. In Fig.~\ref{fig:q2cfts}, we show the spectra of the scaling dimensions at the fifth RG step using Eq.~(\ref{eq:BCFT_tower}). The dashed lines and small figures near the data plots represent the exact values and degeneracies respectively~\cite{Francesco_CFT}. We adopt three boundary conditions discussed above. In Fig.~\ref{fig:q2cfts} (a) the result is shown for two free boundary conditions, where the surface coupling in the edges are both $K_s=K$. Figure~\ref{fig:q2cfts} (b) is obtained from two fixed boundary conditions. In Fig.~\ref{fig:q2cfts} (c), the result with mixed boundary condition is shown, where the one edge is fixed while the other is free. We note that the lowest conformal dimension $h_0$ for the mixed boundary condition is set to $1/16$ in contrast to the other boundary conditions. Therefore the conformal spectra in Fig.~\ref{fig:q2cfts} (a), (b), and (c) correspond to the operator contents of Eq.~(\ref{eq:free-free}), (\ref{eq:fix-fix}), and (\ref{eq:free-fix}) respectively. Comparing with these exact values, BTRG gives the correct conformal data depending on the various boundary conditions with good precision. In the insets of Fig.~\ref{fig:q2cfts}, we show the flow of the central charge and conformal dimensions up to 20 RG steps, where the central charge is shown as the red dashed line. As typically observed in TRG-type calculation of the scaling dimensions, after reaching the right values they gradually collapse starting from higher scaling dimensions. In the present case, they converge at around fifth step, and then the tower is collapsing from larger conformal dimensions. To make the flow more stable and achieve higher accuracy, we would need to eliminate the short correlation loop.

\begin{table*}[htb]
\caption{The flow of central charge obtained from boundary tensors with $\chi=72$. For comparison, we also show that computed from the bulk tensors assuming the periodic boundary condition. Notice that the exact value is $c=0.5$.\label{tab:centralcharge}}
\begin{ruledtabular}
\begin{tabular}{ccccccccccccc}
  RG step & 1 & 2 & 3 & 4 & 5 & 6 & 7 & 8 & 9 & 10 & 11 & 12\\\colrule
  boundary& 1.4513 & 0.9776 & 0.6920 & 0.5864 & 0.5411 & 0.5208 & 0.5123 & 0.5101 & 0.5086 & 0.5032 & 0.4954 & 0.4889\\
  bulk& 0.5641 & 0.5188 & 0.5038 & 0.5009 & 0.5002 & 0.4999 & 0.4994 & 0.4984 & 0.4970 & 0.4948 & 0.4920 & 0.4877\\
\end{tabular}
\end{ruledtabular}
\end{table*}
In Table~\ref{tab:centralcharge}, we show the obtained value of central charge at each RG step for two free boundary conditions. In addition to the results from the boundary tensors, those from the bulk tensors with the periodic boundary condition are also denoted together. While the value from the bulk tensors is convergent around the fifth step to the exact value $c=0.5$ with good precision, the obtained value from the boundary tensors more slowly converges with worse accuracy. This suggests when we estimate the central charge of an unknown model by BTRG, the central charge should be also computed from bulk tensors to be compared with that from the boundary ones, not to draw a wrong conclusion.
%

\section{Conclusion\label{sec:conclusion}}

In this paper, we proposed a new method of investigating the boundary property of the statistical system. We generalized the HOTRG algorithm to make it possible to deal with the system with open boundaries, and simulated the two-dimensional Ising model in the annulus geometry. The spontaneous magnetization at surface and bulk computed by the impurity method gives quantitatively correct results comparing with the exact calculation. In addition, we analyzed the fixed point feature of the boundary tensors, which correctly represented the degeneracy of each of the disordered phase and the $Z_2$ symmetry broken phase. At the critical point, we defined the scale invariant boundary tensor, from which we successfully extracted the information of the conformal data described by the $\mathcal{M}_{4,3}$ minimal BCFT of the annulus with various boundary states. Therefore, BTRG is another numerical method to investigate the BCFT of lattice models than the exact diagonalization~\cite{Gehlen_1986}, DMRG~\cite{PRB.88.075112,Lauchli2013}, and the entanglement renormalization~\cite{PRB.82.161107}.
Because it is straightforeward to extend BTRG for the higher dimension, it would also be useful to investigate the surface critical behavior of three dimensional systems, although precise BCFT analysis is difficult.

However, since we do not eliminate the short correlation loop remaining in the network, the flow of the conformal data is unstable and the precision is not so good. It is expected that combining with such an algorithm would make the results better. If using a symmetry broken boundary condition, such as the Cardy states in Eq.~(\ref{eq:Cardy+}) or (\ref{eq:Cardy-}), since we cannot utilize the $Z_2$ symmetric tensor it would be more important to eliminate the short loops to achieve the good accuracy with smaller bond dimensions.
%

\appendix

\section{Construction of the projectors\label{sec:projector}}

In this appendix, we discuss how to obtain the proper projectors in general situation to renormalize the four tensors forming a plaquette into two tensors, as
\begin{equation}
\begin{minipage}{1.5truecm}
      \centering
      \includegraphics[width=1.5truecm,clip]{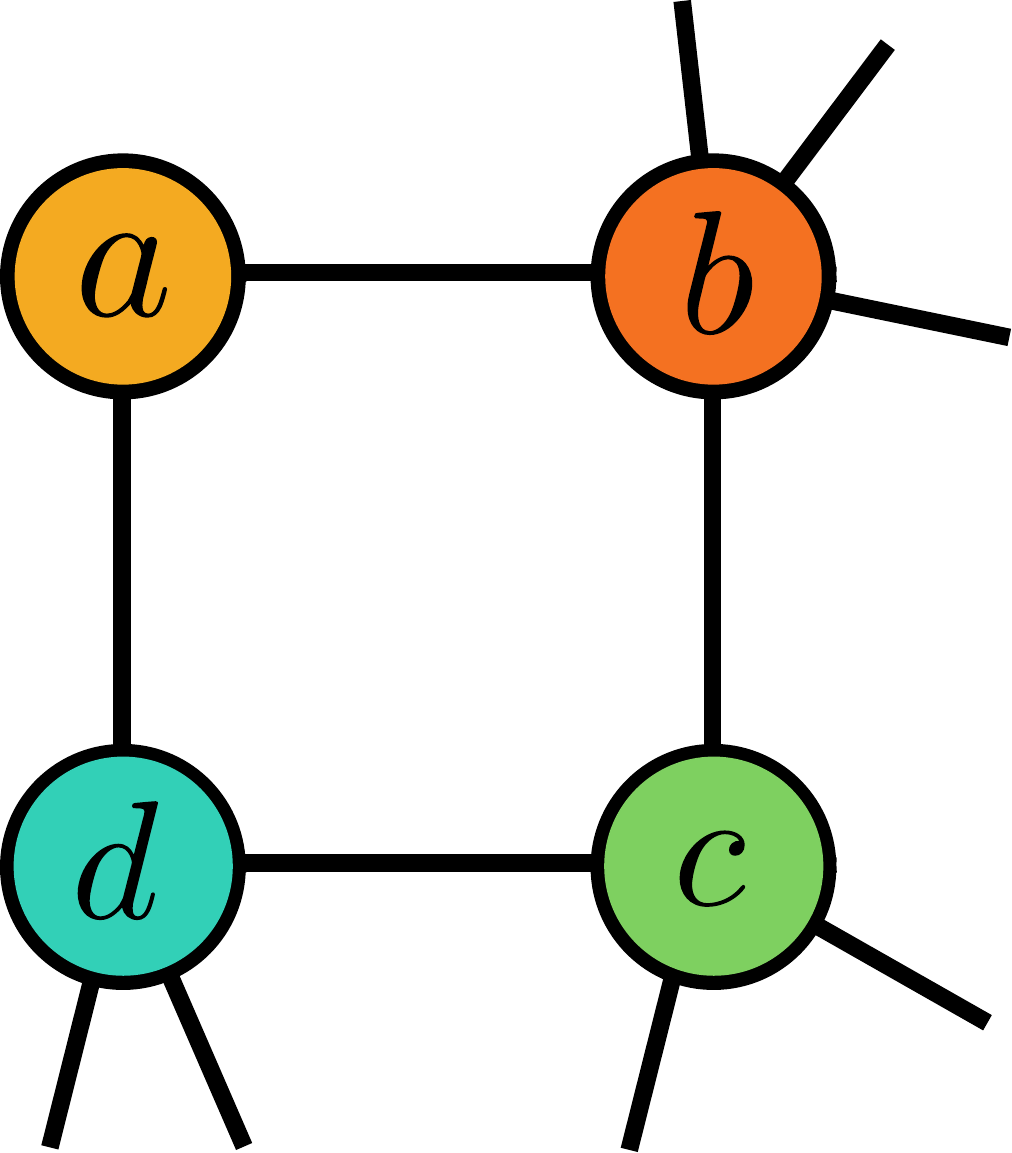}
\end{minipage}
\rightarrow
\begin{minipage}{1.6truecm}
      \centering
      \includegraphics[width=1.6truecm,clip]{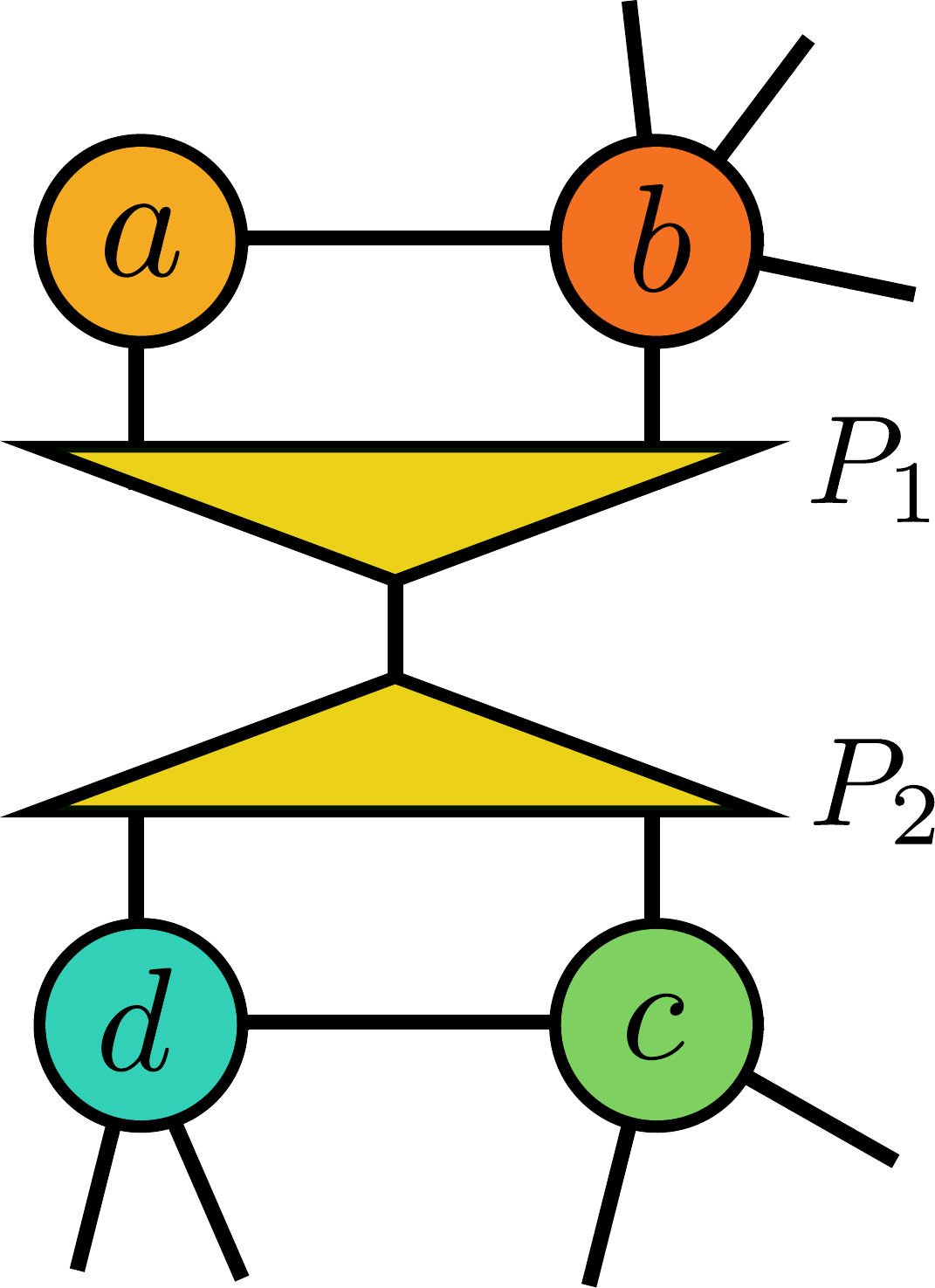}
\end{minipage}
\rightarrow
\begin{minipage}{0.8truecm}
      \centering
      \includegraphics[width=0.8truecm,clip]{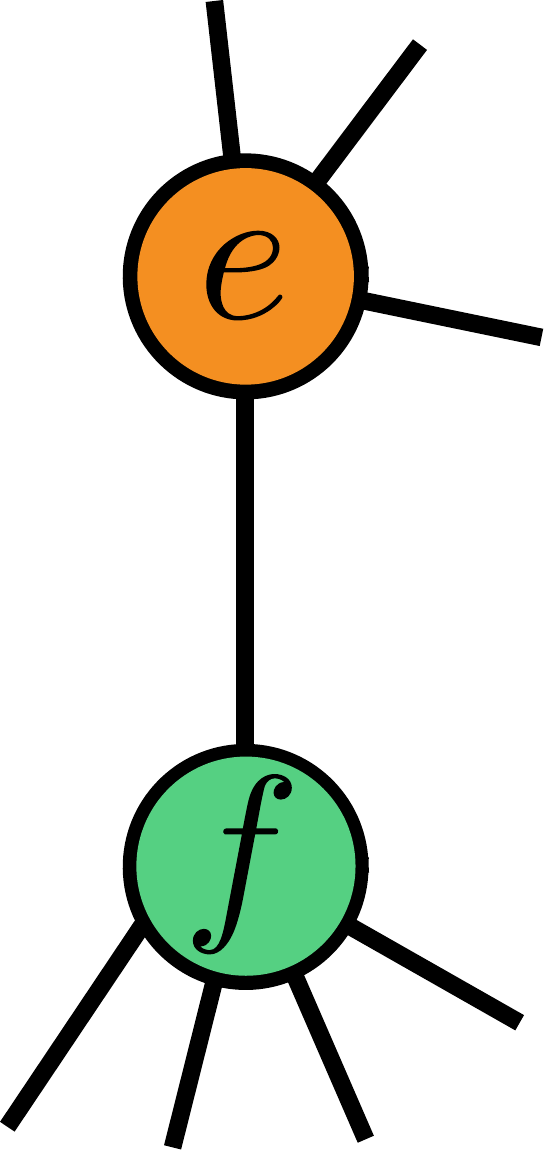}
\end{minipage}
.
\label{eq:RGprocess}
\end{equation}
Such projectors $P_1$ and $P_2$ can be determined so as to minimize the norm
\begin{equation}
  \mathcal{C}=\left|
  \begin{minipage}{1.5truecm}
      \centering
      \includegraphics[width=1.5truecm,clip]{plaquette_gen.pdf}
\end{minipage}
-
\begin{minipage}{1.6truecm}
      \centering
      \includegraphics[width=1.6truecm,clip]{plaquette_gen-prj.pdf}
\end{minipage}
\right|^2.
\label{eq:costfunc_gen}
\end{equation}
In order to take unnecessary bonds aside, we consider the following QR decomposition~\cite{Wang2011,PRL.113.046402,Yang2017}:
\begin{eqnarray}
  \begin{minipage}{1.5truecm}
      \centering
      \includegraphics[width=1.5truecm,clip]{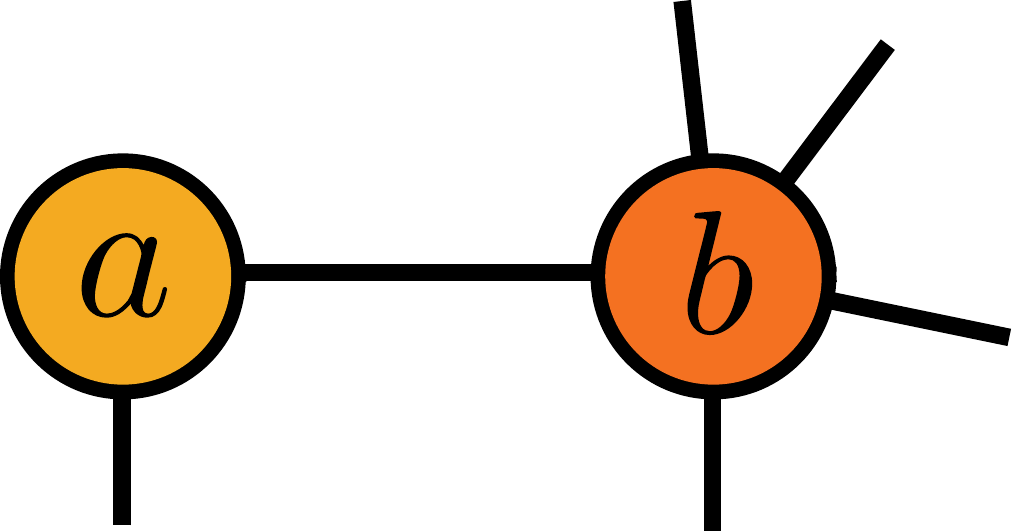}
\end{minipage}
&=&
\begin{minipage}{1.6truecm}
      \centering
      \includegraphics[width=1.6truecm,clip]{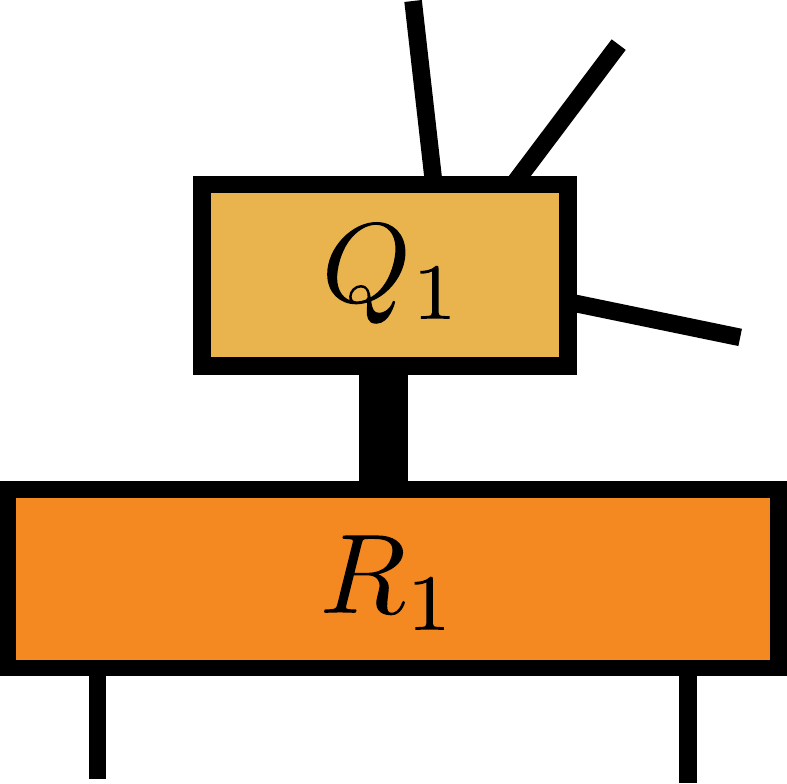}
\end{minipage}
\label{eq:qrdecomposition1}\\
  \begin{minipage}{1.5truecm}
      \centering
      \includegraphics[width=1.5truecm,clip]{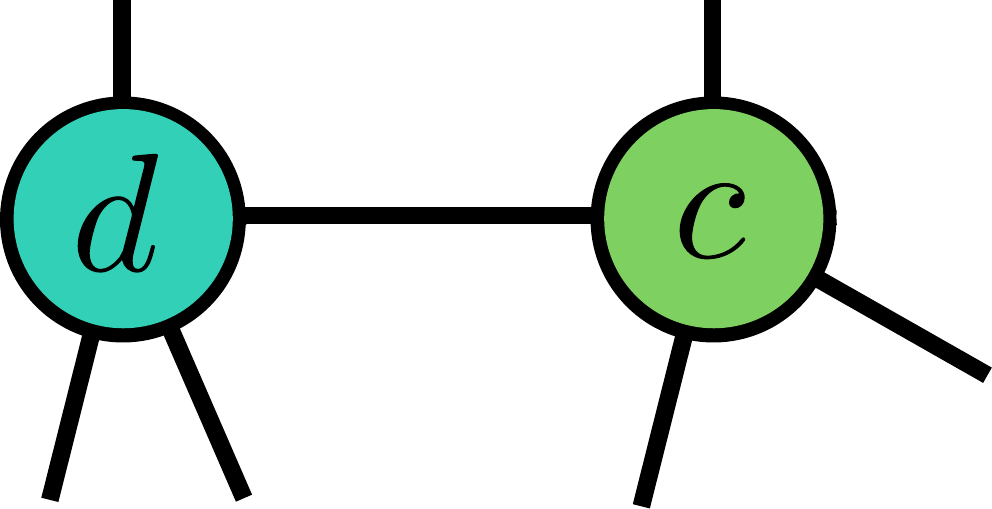}
\end{minipage}
&=&
\begin{minipage}{1.6truecm}
      \centering
      \includegraphics[width=1.6truecm,clip]{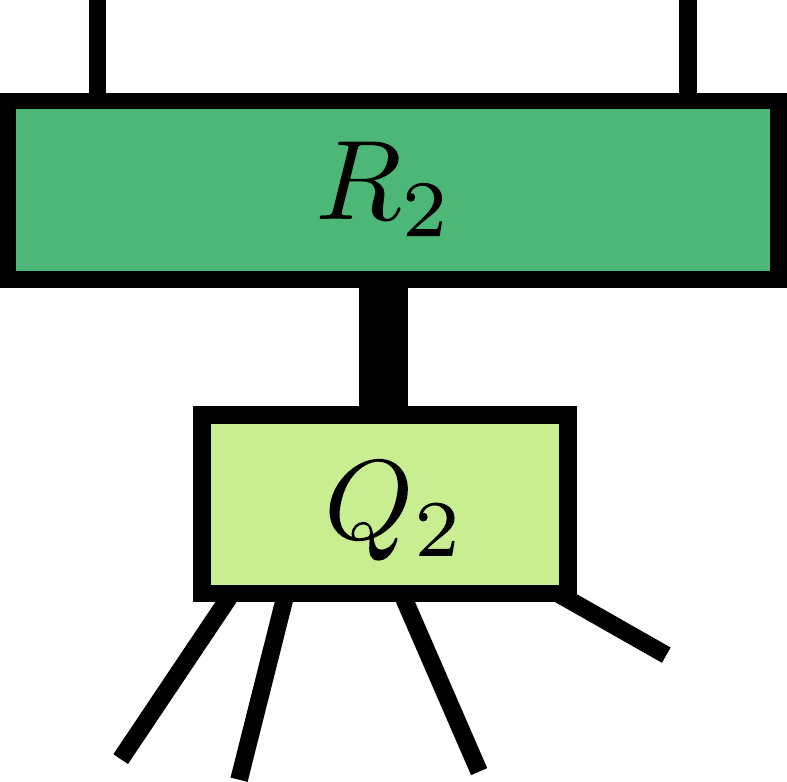}
\end{minipage}.
\label{eq:qrdecomposition2}
\end{eqnarray}
However, notice that we do not have to perform this QR decomposition, because for example
\begin{equation}
  \begin{minipage}{1.5truecm}
      \centering
      \includegraphics[width=1.5truecm,clip]{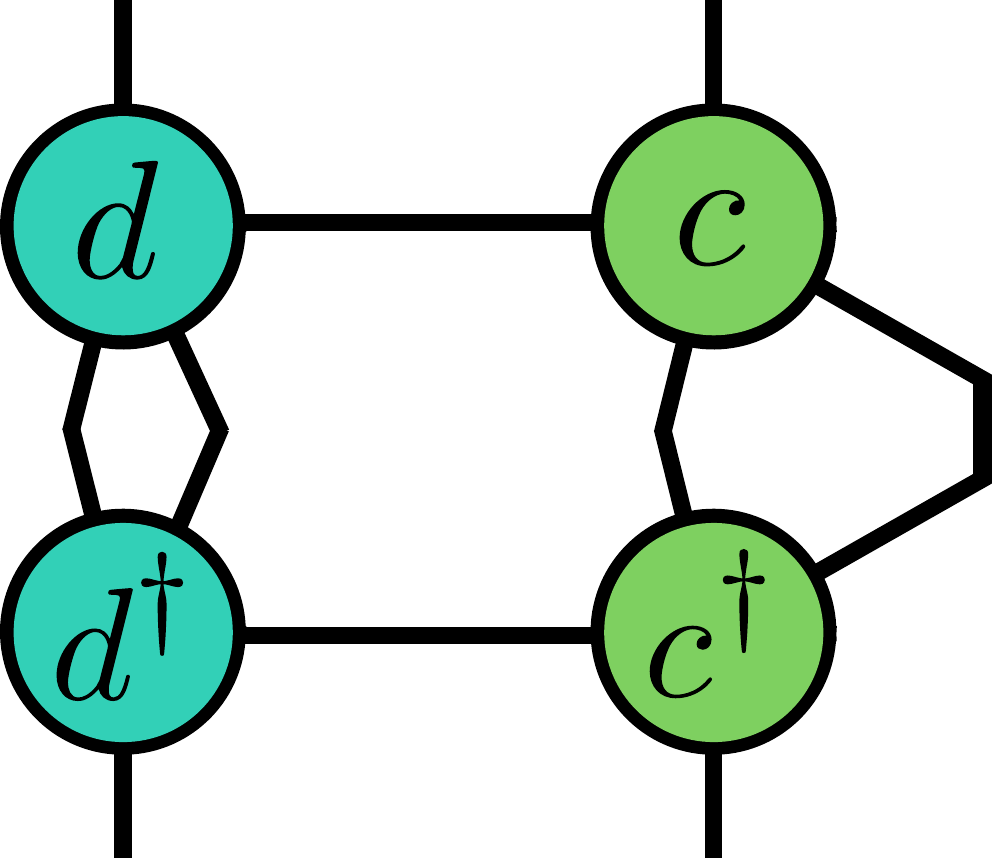}
\end{minipage}
=
\begin{minipage}{1.5truecm}
      \centering
      \includegraphics[width=1.5truecm,clip]{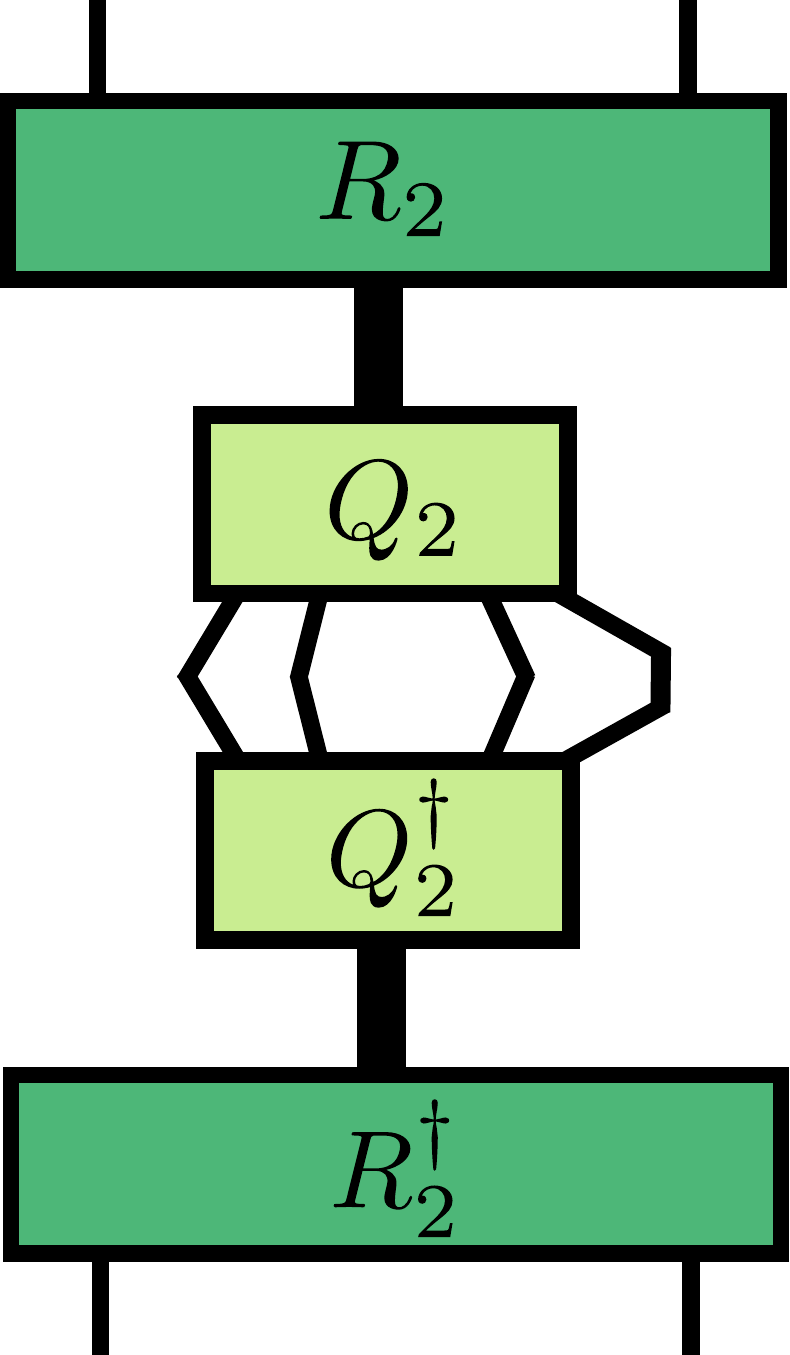}
\end{minipage}
=
\begin{minipage}{1.5truecm}
      \centering
      \includegraphics[width=1.5truecm,clip]{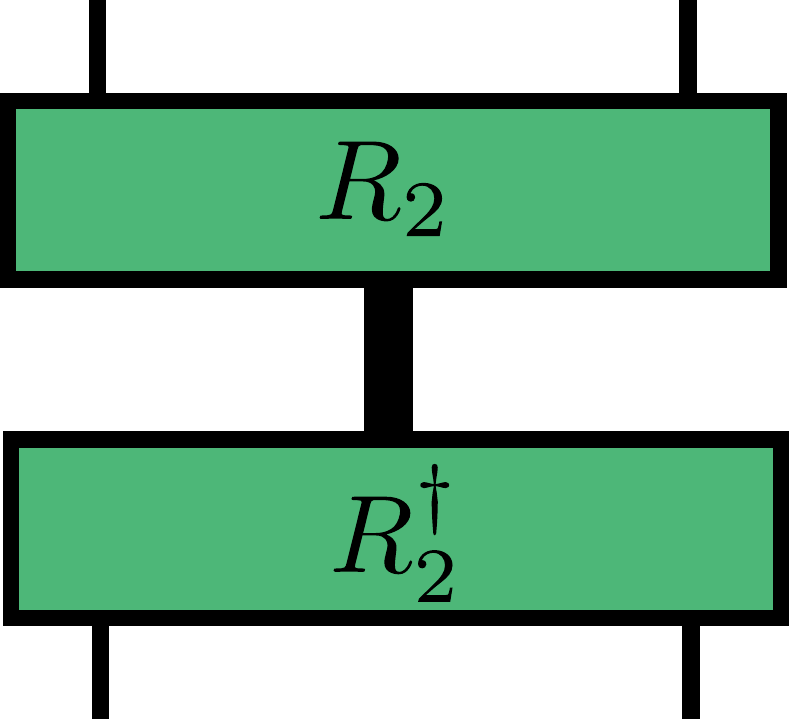}
\end{minipage},
\label{eq:obtainR}
\end{equation}
where $Q_2Q_2^{\dagger}={\boldsymbol 1}$ by definition. Namely, we can compute $R_2$ by the singular value decomposition (SVD) (or eigenvalue decomposition) of the left-side tensor in Eq.~(\ref{eq:obtainR}). For two-dimensional square lattice, while the computational cost for Eq.~(\ref{eq:qrdecomposition1}) and (\ref{eq:qrdecomposition2}) is $O(\chi^8)$, Eq.~(\ref{eq:obtainR}) reduces the cost to $O(\chi^6)$.

To determine the projectors, contract $R_1$ and $R_2$ and then perform the singular value decomposition for it:
\begin{equation}
  \begin{minipage}{1.5truecm}
      \centering
      \includegraphics[width=1.5truecm,clip]{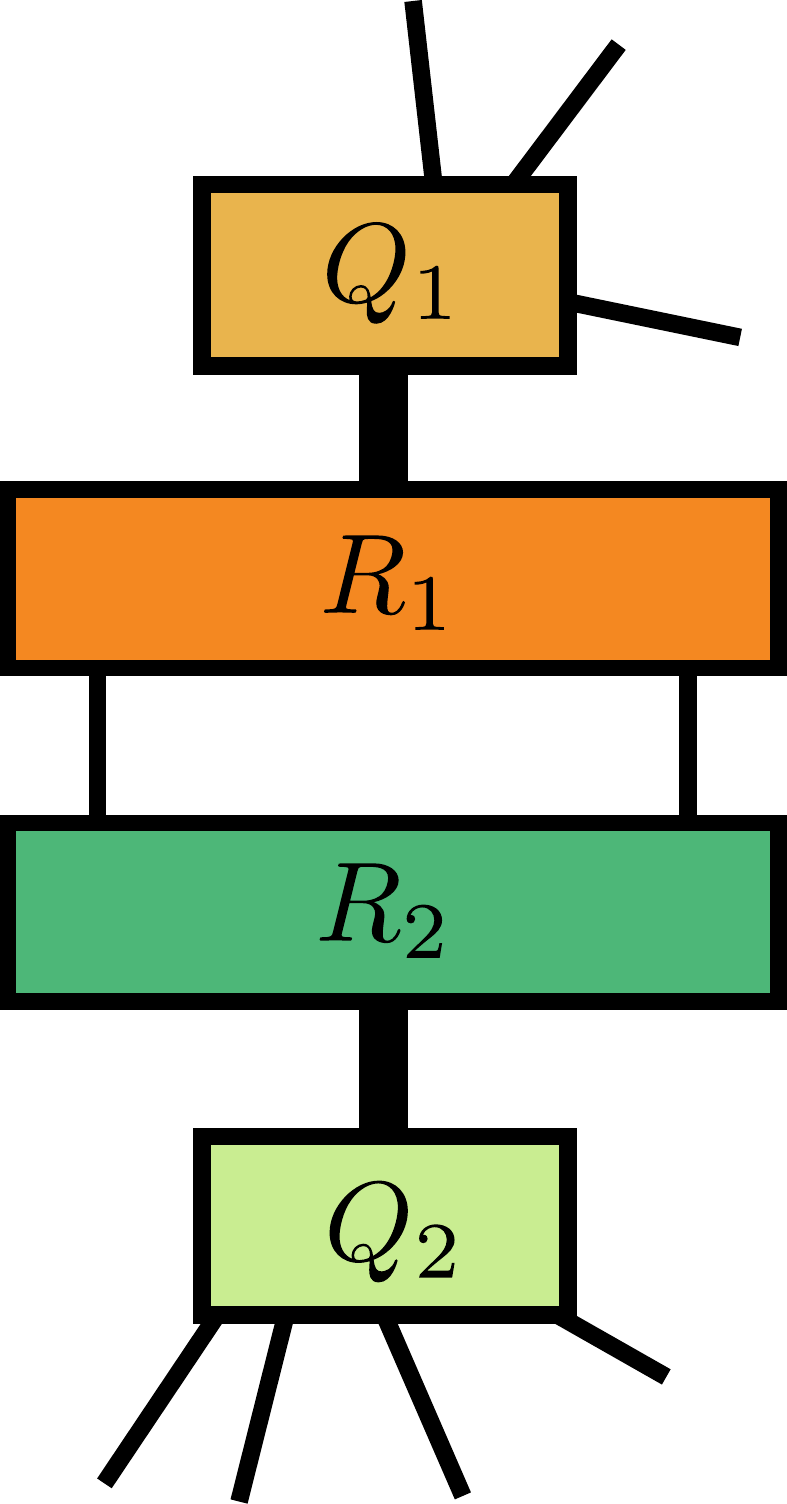}
\end{minipage}
\simeq
\begin{minipage}{1.5truecm}
      \centering
      \includegraphics[width=1.5truecm,clip]{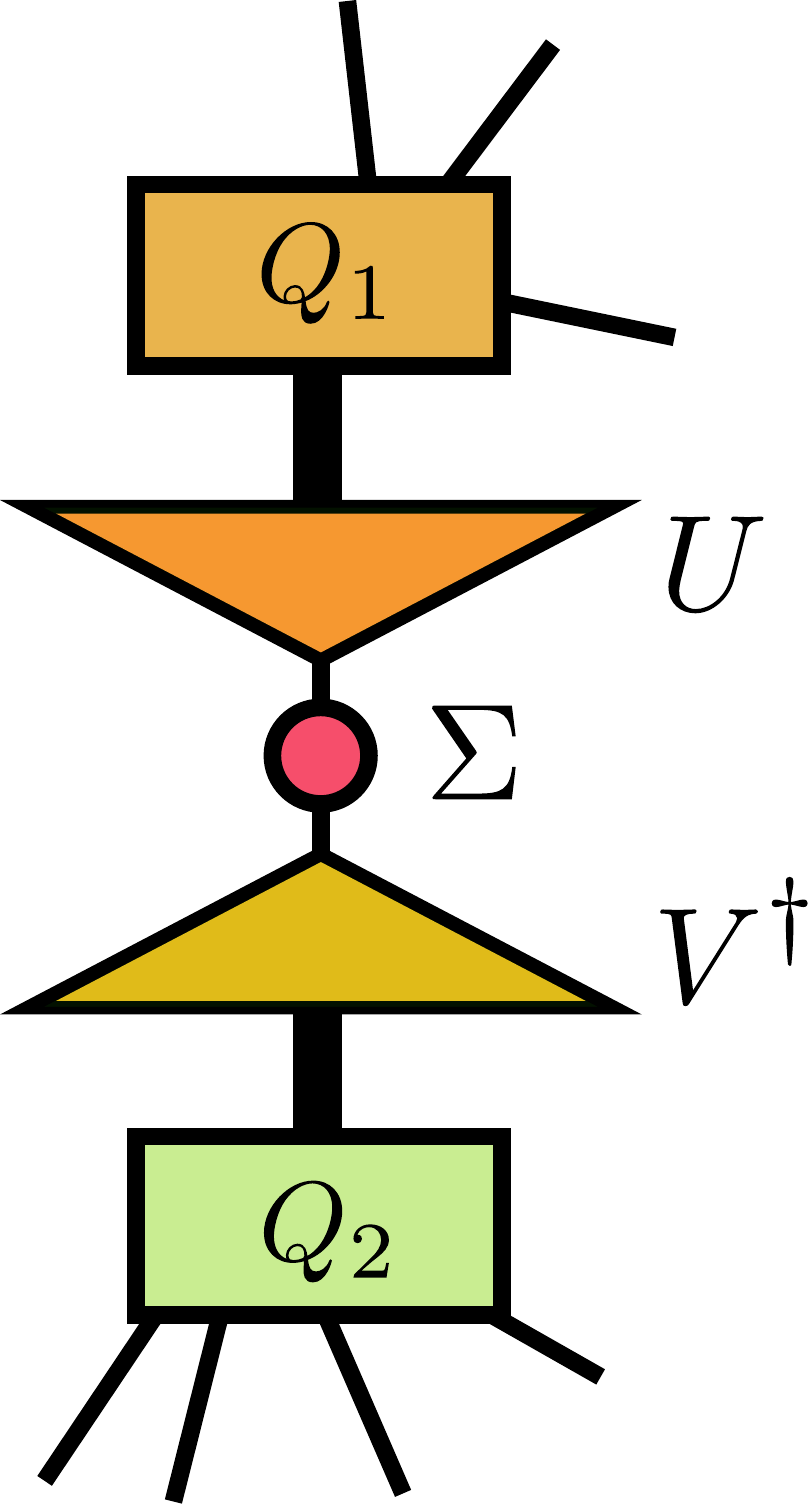}
\end{minipage}.
\label{eq:svdR1R2}
\end{equation}
Note that $\Sigma$ is the truncated singular value vector. Comparing the right hand side of Eq.~(\ref{eq:svdR1R2}) with the middle of Eq.~(\ref{eq:RGprocess}),
\begin{equation}
  \begin{minipage}{1.5truecm}
      \centering
      \includegraphics[width=1.5truecm,clip]{plaquette_gen-prj.pdf}
\end{minipage}
=
  \begin{minipage}{1.9truecm}
      \centering
      \includegraphics[width=1.9truecm,clip]{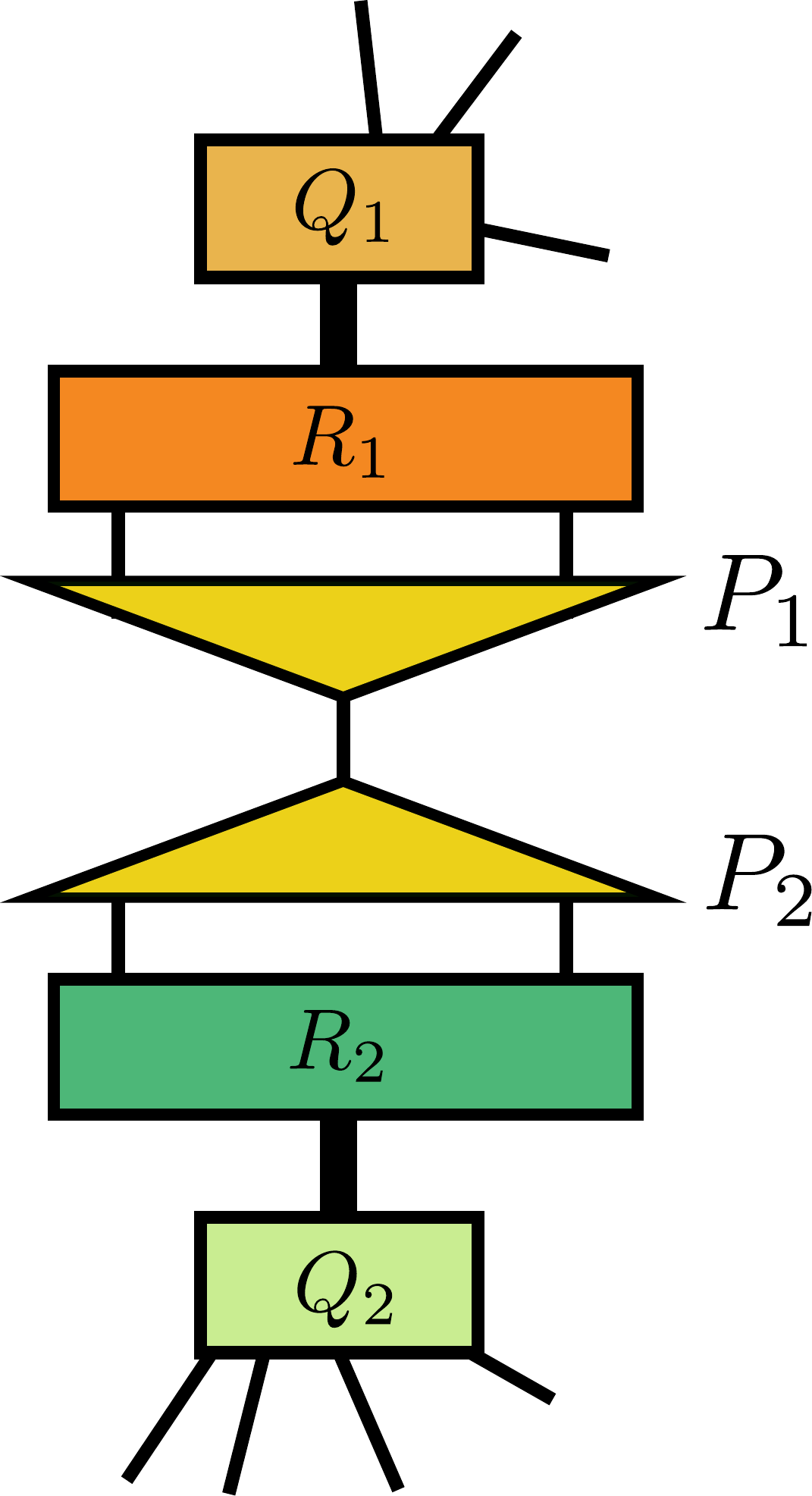}
\end{minipage}
\simeq
\begin{minipage}{1.5truecm}
      \centering
      \includegraphics[width=1.5truecm,clip]{svdR1R2.pdf}
\end{minipage},
\label{eq:compare_prj}
\end{equation}
amounts to $R_1P_1=U\sqrt{\Sigma}$ and $P_2R_2=\sqrt{\Sigma}V^{\dagger}$. We can show that these projectors minimizes the cost function Eq.~(\ref{eq:costfunc_gen}) in the sense of the Frobenius norm. To avoid computing the inverse of $R_1$ and $R_2$, we can make use of the result of SVD, $R_1R_2\simeq U\Sigma V^{\dagger}$, which yields
\begin{eqnarray}
  U\sqrt{\Sigma}&\simeq& R_1R_2V\sqrt{\Sigma}^{-1}\\
  \sqrt{\Sigma}V^{\dagger}&\simeq& \sqrt{\Sigma}^{-1}U^{\dagger}R_1R_2.
\label{eq:svdresult}
\end{eqnarray}
Therefore, the projectors are finally
\begin{eqnarray}
\begin{minipage}{1.8truecm}
      \centering
      \includegraphics[width=1.8truecm,clip]{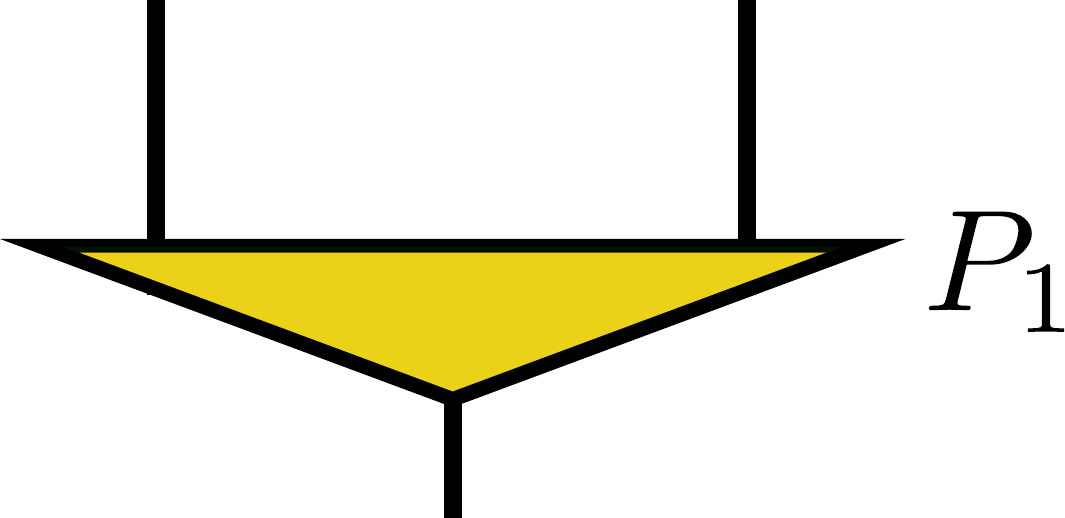}
\end{minipage}
=
\begin{minipage}{1.8truecm}
      \centering
      \includegraphics[width=1.8truecm,clip]{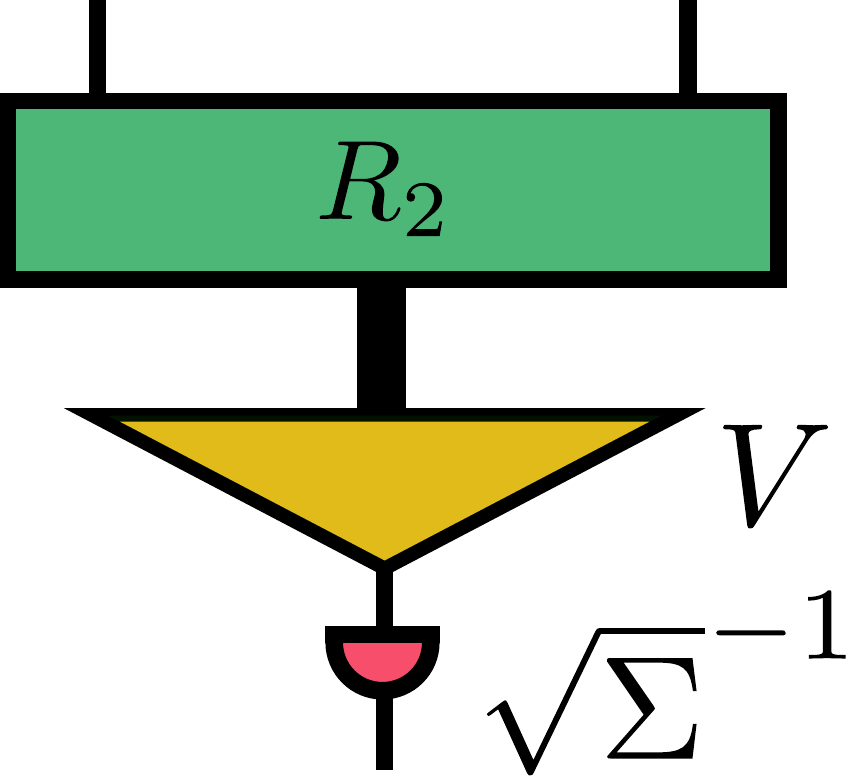}
\end{minipage}\\
\begin{minipage}{1.8truecm}
      \centering
      \includegraphics[width=1.8truecm,clip]{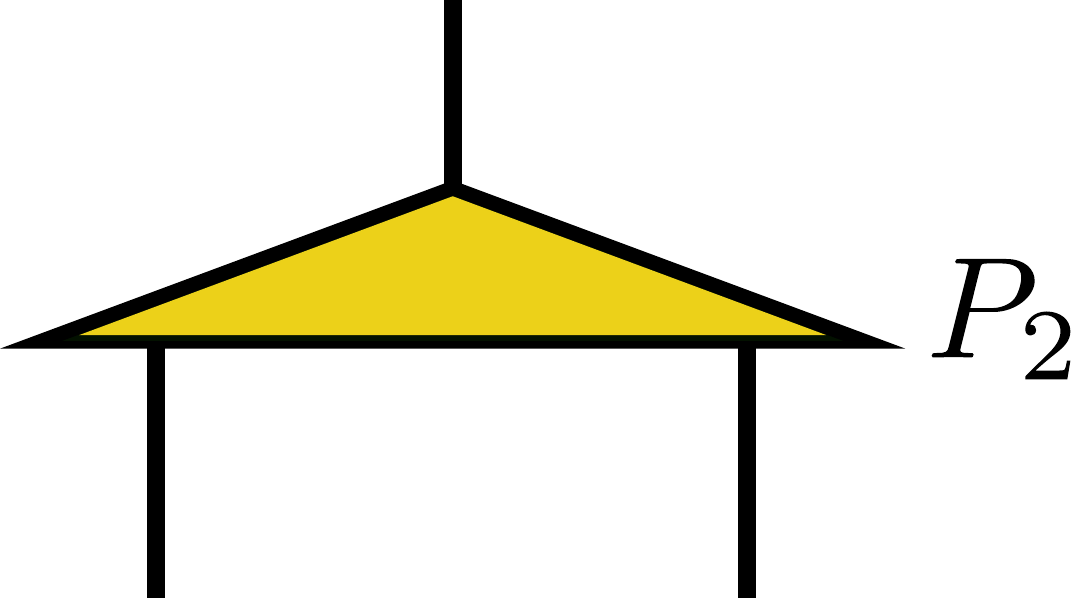}
\end{minipage}
=
\begin{minipage}{1.8truecm}
      \centering
      \includegraphics[width=1.8truecm,clip]{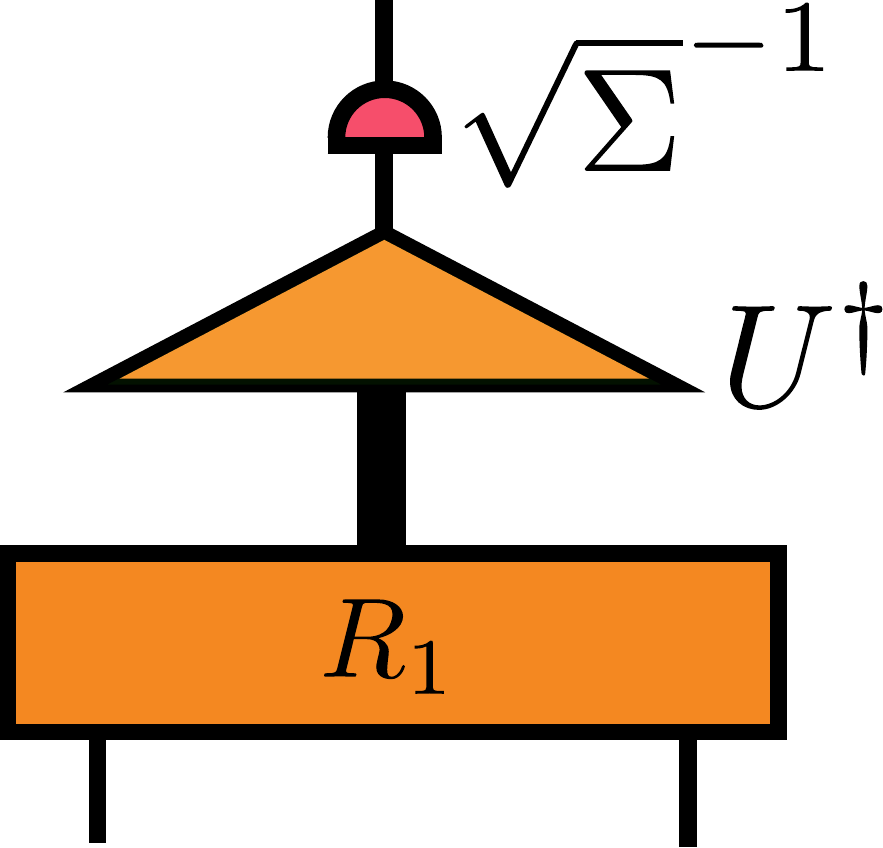}
\end{minipage}
\label{eq:projector}
\end{eqnarray}
The overall computational cost for creating projectors requires $O(\chi^6)$, which is the same as that of the higher-order SVD in HOTRG algorithm for two-dimensional square lattice. Actually, for the isotropic tensor the projector created for the bulk tensors are the same as the one for HOTRG. In this sense this way of constructing the projector is the natural generalization of HOTRG algorithm.

\section{Calculation of the scale invariant tensor\label{sec:invtensor}}

In this appendix, we describe how to obtain the scale invariant tensor, which is necessary to compute the conformal data as explained in Sec.~\ref{subsec:fptensor}. Let us suppose we hold the bulk tensor $A^{(i)}$ and boundary tensors $B^{(i)}_1$ and $B^{(i)}_2$ in the $i$-th RG step, the number of which are $N_a^{(i)}$, $N_b^{(i)}$ and also $N_b^{(i)}$ respectively. The partition function can be symbolically described as
\begin{equation}
  Z = \tTr \otimes \left[{B^{(i)}_1}^{N_b^{(i)}} {A^{(i)}}^{N_a^{(i)}} {B^{(i)}_2}^{N_b^{(i)}}\right].
  \label{eq:Z}
\end{equation}
To avoid the overflow of exponentially growing partition function, in practice we normalize the tensors. Here let us define the normalized tensors
\begin{equation}
  a^{(i)}=A^{(i)}/\Gamma_a^{(i)},\ b_1^{(i)}=B_1^{(i)}/\Gamma_b^{(i)},\ b_2^{(i)}=B_2^{(i)}/\Gamma_b^{(i)},
  \label{eq:normalized_tensor}
\end{equation}
and impose the following normalization:
\begin{equation}
\begin{minipage}{1.1truecm}
      \centering
      \includegraphics[width=1.1truecm,clip]{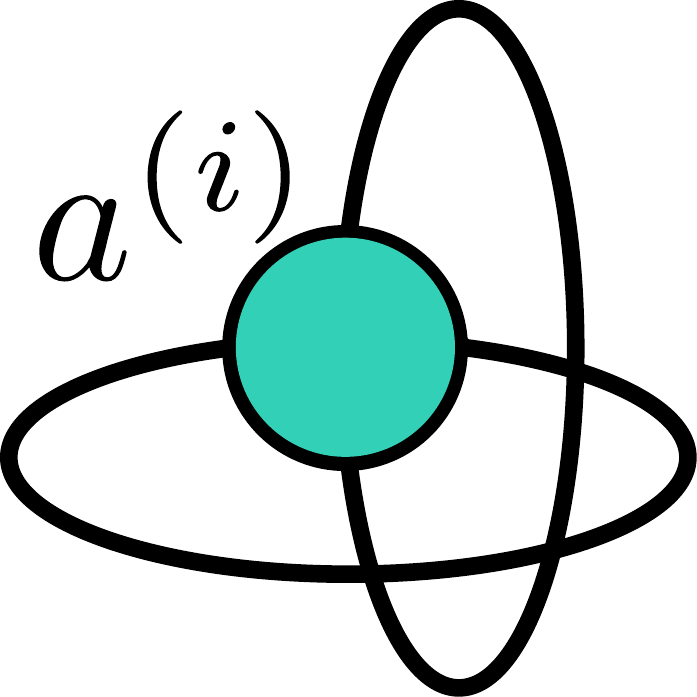}
\end{minipage}=1,\ \ 
\begin{minipage}{0.9truecm}
      \centering
      \includegraphics[width=0.9truecm,clip]{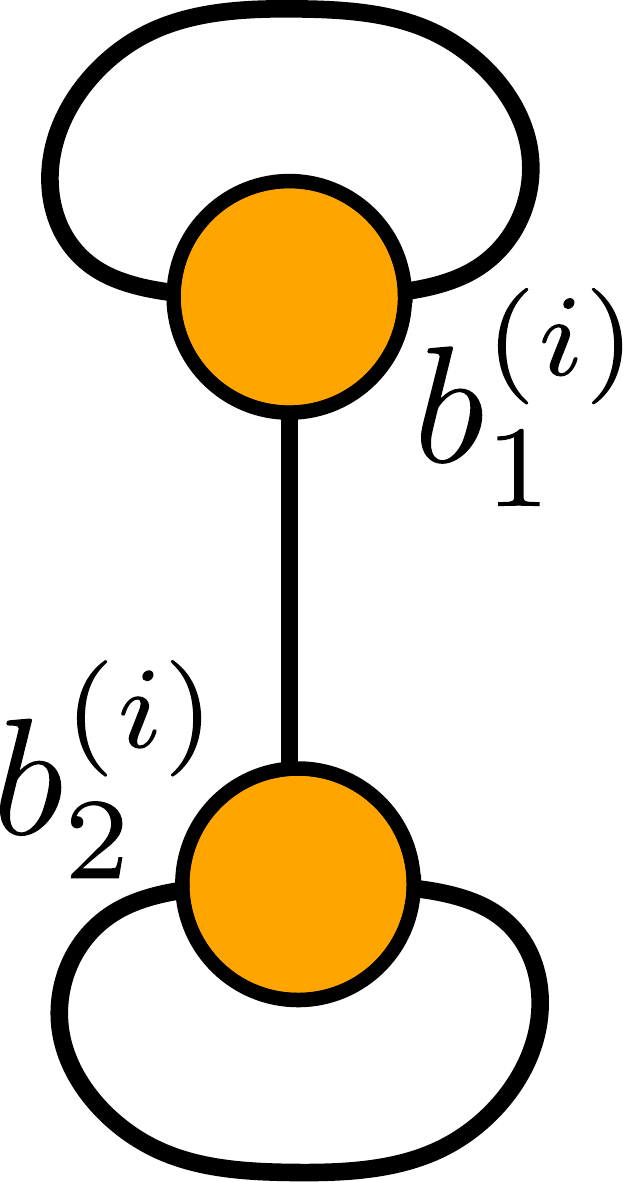}
\end{minipage}=1.
  \label{eq:condition_normalize}
\end{equation}
Using them, the partition function is
\begin{equation}
  Z = {\Gamma_a^{(i)}}^{N_a^{(i)}}{\Gamma_b^{(i)}}^{2N_b^{(i)}}
  \tTr \otimes \left[{b^{(i)}_1}^{N_b^{(i)}} {a^{(i)}}^{N_a^{(i)}} {b^{(i)}_2}^{N_b^{(i)}}\right].
  \label{eq:Z_normalize}
\end{equation}
Now let us define scale invariant tensors
\begin{eqnarray}
  a^{(i)}_{\mathrm{inv}} &=& \gamma_a^{-1}a^{(i)},
  \label{eq:inv_tensora}\\
  b^{(i)}_{1\mathrm{inv}} &=& \gamma_1^{-1}b^{(i)}_1 \ \ \mathrm{and}\ \ b^{(i)}_{2\mathrm{inv}} = \gamma_2^{-1}b^{(i)}_2,
  \label{eq:inv_tensorb}
\end{eqnarray}
so as to satisfy the following condition
\begin{widetext}
\begin{equation}
  \tTr \otimes \left[{b^{(i)}_{1\mathrm{inv}}}^{N_b^{(i)}} {a^{(i)}_{\mathrm{inv}}}^{N_a^{(i)}} {b^{(i)}_{2\mathrm{inv}}}^{N_b^{(i)}}\right]=
  \tTr \otimes \left[{b^{(i+1)}_{1\mathrm{inv}}}^{N_b^{(i+1)}} {a^{(i+1)}_{\mathrm{inv}}}^{N_a^{(i+1)}} {b^{(i+1)}_{2\mathrm{inv}}}^{N_b^{(i+1)}}\right].
  \label{eq:inv_condition}
\end{equation}
Two equations of Eq.~(\ref{eq:Z_normalize}) and Eq.~(\ref{eq:inv_condition}) and invariance of the partition function for RG transformation leads to
\begin{equation}
  {\Gamma_a^{(i)}}^{N_a^{(i)}}{\Gamma_b^{(i)}}^{2N_b^{(i)}}\gamma_a^{N_a^{(i)}}(\gamma_1\gamma_2)^{N_b^{(i)}}=
  {\Gamma_a^{(i+1)}}^{N_a^{(i+1)}}{\Gamma_b^{(i+1)}}^{2N_b^{(i+1)}}\gamma_a^{N_a^{(i+1)}}(\gamma_1\gamma_2)^{N_b^{(i+1)}}.
  \label{eq:recursion_relation1}
\end{equation}
\end{widetext}
Here, our renormalization procedure described in Fig.~\ref{fig:RGstep} (c) gives
\begin{eqnarray}
  2N_b^{(i+1)} &=& N_b^{(i)}\\
  4N_a^{(i+1)} &=& N_a^{(i)}-2N_b^{(i)},
  \label{eq:recursion_relation2}
\end{eqnarray}
which simplifies Eq.~(\ref{eq:recursion_relation1}) by assuming that in $(i+1)$th step only the boundary tensors remain (i.e., $N_a^{(i+1)}=0$):
\begin{equation}
  \gamma_1\gamma_2 = \left(\gamma_a\Gamma_a^{(i)}\right)^{-4}\left(\frac{\Gamma_b^{(i+1)}}{{\Gamma_b^{(i)}}^2}\right)^2.
  \label{eq:factor_boundary1}
\end{equation}
It allows us to determine $\gamma_a$ assuming the bulk tensor $a$ is scale invariant for the bulk RG,
\begin{equation}
\begin{minipage}{1.75truecm}
      \centering
      \includegraphics[width=1.75truecm,clip]{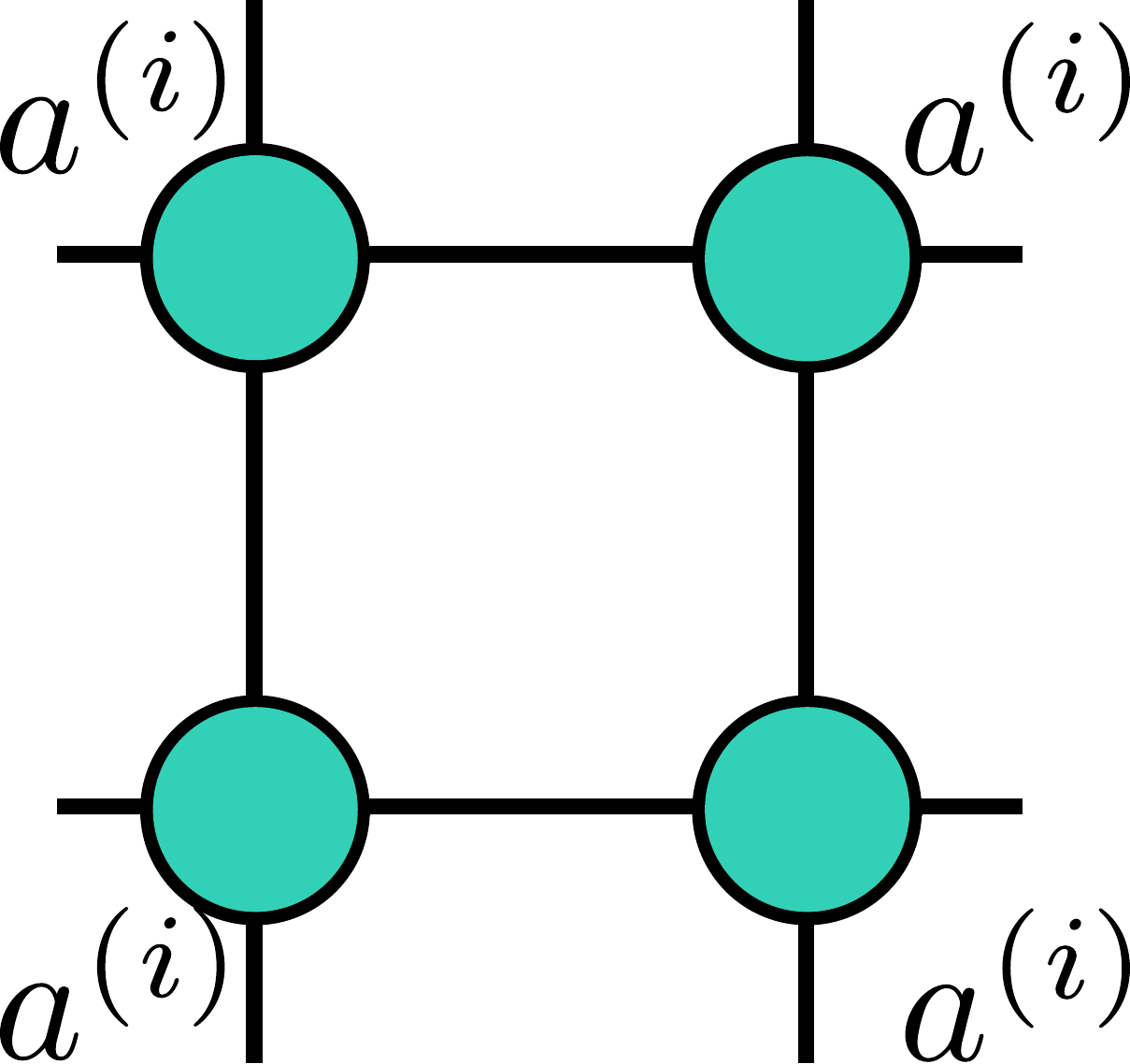}
\end{minipage}
\rightarrow
\begin{minipage}{1.2truecm}
      \centering
      \includegraphics[width=1.2truecm,clip]{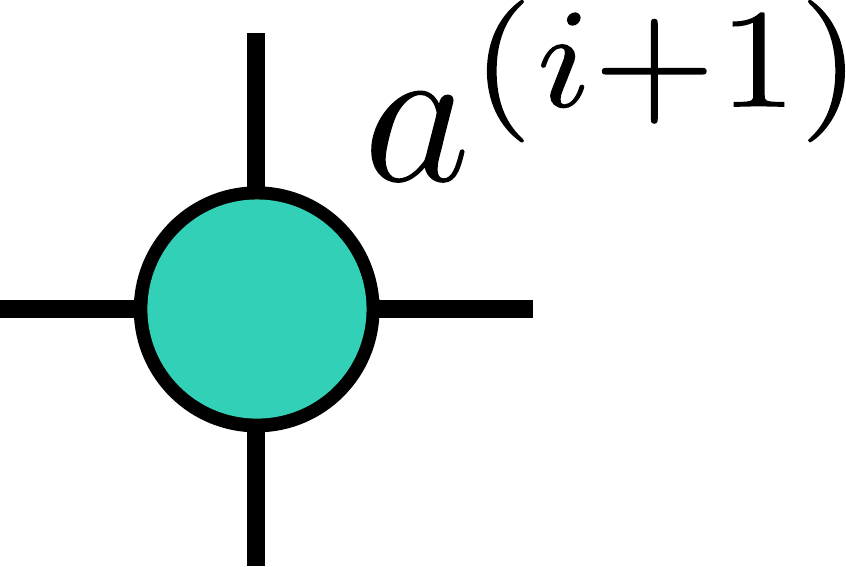}
\end{minipage}
,
\end{equation}
which is, as is also discussed in the appendix of Ref.~\onlinecite{Gu2009},
\begin{equation}
  \gamma_a=\left(\frac{\Gamma_a^{(i+1)}}{{\Gamma_a^{(i)}}^4}\right)^{\frac{1}{3}}.
  \label{eq:factor_bulk}
\end{equation}
Finally, we obtain
\begin{equation}
  \gamma_1\gamma_2 = \left(\frac{\Gamma_a^{(i+1)}}{\Gamma_a^{(i)}}\right)^{-\frac{4}{3}}\left(\frac{\Gamma_b^{(i+1)}}{{\Gamma_b^{(i)}}^2}\right)^2.
  \label{eq:factor_boundary2}
\end{equation}
Note that it is not necessary to know the $\gamma_1$ and $\gamma_2$ separately because we always use both of $b_1$ and $b_2$ to construct a transfer matrix, such as Eq.~(\ref{eq:trfmat_bb}).

The relation between $\Gamma^{(i+1)}$ and $\Gamma^{(i)}$ is
\begin{eqnarray}
  \ln\Gamma^{(i+1)}_b-2\ln\Gamma^{(i)}_b&=&2\ln\Gamma^{(i)}_a+\ln\sqrt{
\begin{minipage}{1.65truecm}
      \centering
      \includegraphics[width=1.65truecm,clip]{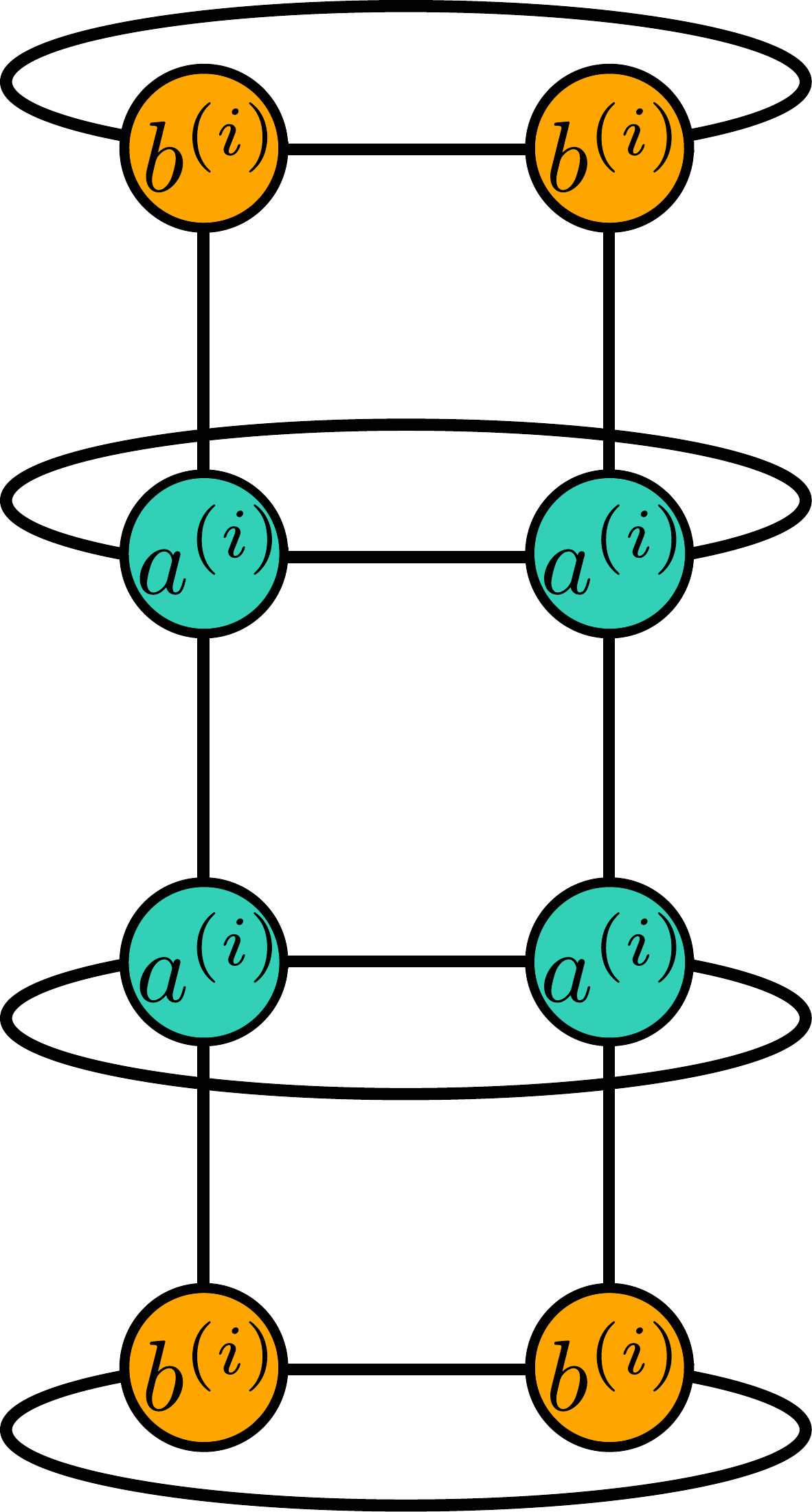}
\end{minipage}
  }\\
  \ln\Gamma^{(i+1)}_a-4\ln\Gamma^{(i)}_a&=&\ln\left[
\begin{minipage}{1.65truecm}
      \centering
      \includegraphics[width=1.65truecm,clip]{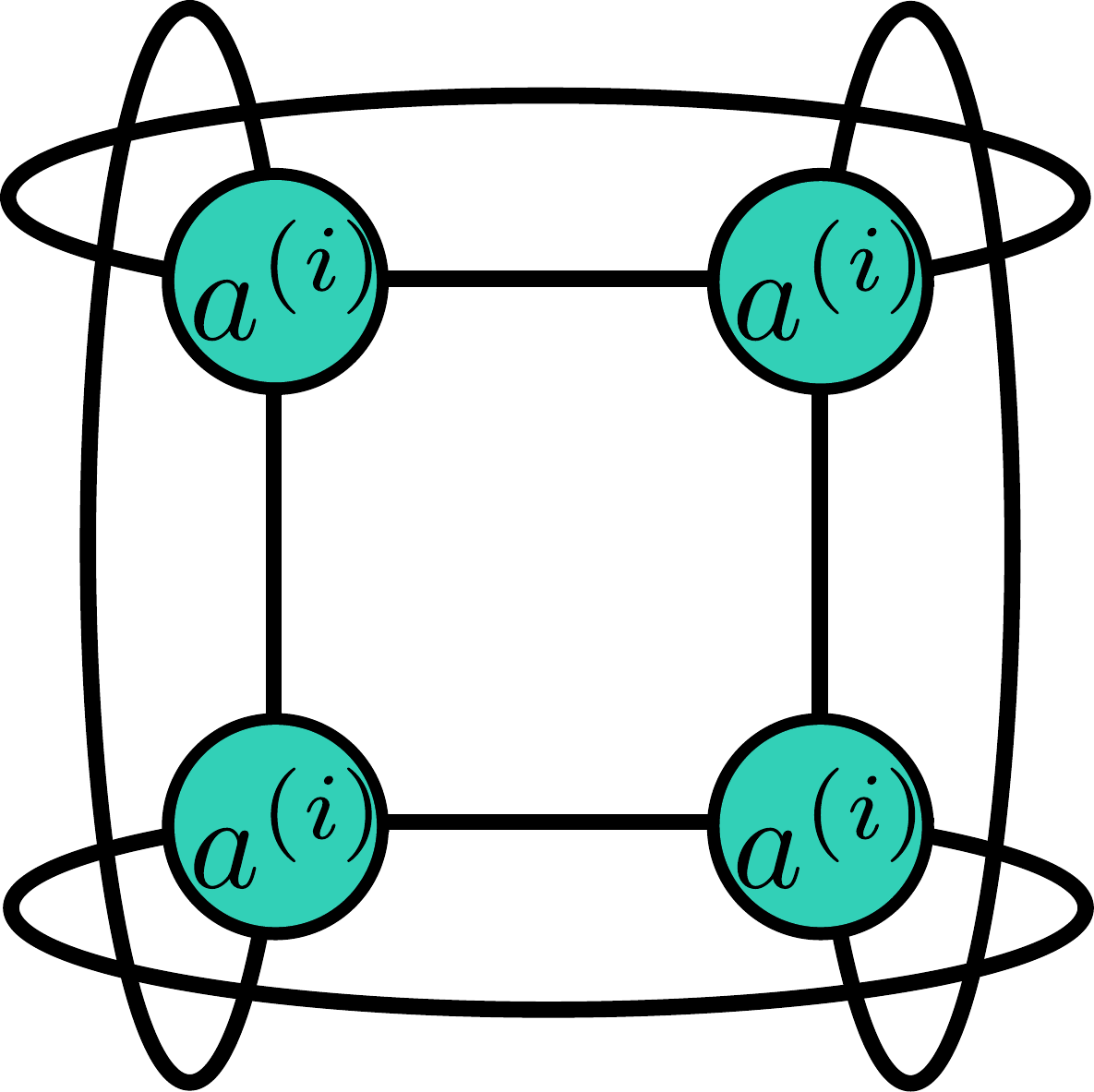}
\end{minipage}\right]
  .
\end{eqnarray}
Substituting them in Eq.~(\ref{eq:factor_boundary2}) yields
\begin{equation}
  \gamma_1\gamma_2=\left[
\begin{minipage}{1.65truecm}
      \centering
      \includegraphics[width=1.65truecm,clip]{plaquette_tra.pdf}
\end{minipage}\right]^{-\frac{4}{3}}\left[
\begin{minipage}{1.65truecm}
      \centering
      \includegraphics[width=1.65truecm,clip]{plaquette_trab.pdf}
\end{minipage}
  \right].
\end{equation}
Now we are able to calculate the scale invariant tensors for each RG step, according to Eq.~(\ref{eq:inv_tensora}) and Eq.~(\ref{eq:inv_tensorb}).
%


\begin{acknowledgments}
S.I. thanks Yoshiki Fukusumi and Tateki Obori for inspiring discussions and comments, and also is grateful to the support of Program for Leading Graduate Schools (ALPS). N.K.'s work is financially supported by MEXT Grant-in-Aid for Scientific Research (B) (25287097, 19H01809). This research was supported by MEXT as "Exploratory Challenge on Post-K computer" (Frontiers of Basic Science: Challenging the Limits).
\end{acknowledgments}

\bibliography{bibliography}

\end{document}